\documentclass{ws-rv9x6}
\usepackage{ws-rv-van}             
\makeindex
\begin{document}

\renewcommand\theequation{\arabic{equation}}
\renewcommand\thesection{\arabic{section}}

\chapter*{Superembedding approach to superstring in $ AdS_5\times S^5$
superspace \label{ch1} }

\author[Igor A. Bandos]{Igor A. Bandos\footnote{igor\_bandos@ehu.es, bandos@ific.uv.es}}

\address{Ikerbasque, the Basque Science Foundation, and  \\ 
Department of Theoretical Physics and History of Science,
\\ the University of the Basque Country, \\ P.O. Box 644, 48080
Bilbao, Spain
 \\ and \\ Institute for Theoretical Physcs, \\
NSC Kharkov Institute of Physics \& Technology, \\
UA 61108,  Kharkov, Ukraine. }

\begin{abstract}
We review the spinor moving frame formulations and generalized
action principle for super-$p$-branes, describe in detail the
superembedding approach to superstring in general type IIB
supergravity background and present the complete superembedding
description of type IIB superstring in the $AdS_5\times S^5$
superspace.
\end{abstract}

\body

This contribution  is devoted to the memory of Wolfgang Kummer who
untimely left us in 2007. We collaborated with him several years
beginning, in 1996, by studying gravity induced on the worldvolume
of a brane \cite{baku99}; this was one of the pre-Rundall-Sundrum
Brane World scenarios (see also \cite{IB97}). Search for its
supersymmetric generalizations led us to thinking on a new form of
D$p$-brane actions \cite{bakuPLB98} and to studying the
super-D$9$-brane dynamics \cite{abkz}. This line was then continued
by attacking the problem of supersymmetric Lagrangian description of
the interacting superbrane systems \cite{bakuPLB2000} which, in my
opinion, still remains open as far as the commonly accepted
candidate action for coincident Dp--branes\cite{Myers:1999ps} does
not possess neither supersymmetry nor Lorentz symmetry.

Among the main tools in our studies were embedding and
superembedding approaches to bosonic and supersymmetric branes. This
is why I decided to chose for my contribution the present manuscript
containing a review of the superembedding approach and its specific
application for the case of superstring in $AdS_5\times S^5$
superspace (see \cite{Metsaev:1998it,Kallosh:1998qs} for
Green--Schwarz superstring action in this superspace).

Notice that superstring in $AdS_5\times S^5$ superspace  is often
called {\it $AdS_5\times S^5$ superstring} (see
\cite{Arutyunov:2009ga} and refs therein).
However, in our opinion, this name might produce an erroneous
impression that the model is essentially different from the
Green--Schwarz (GS) superstring. Such a confusion might be further
enlarged by an accent which is made in the literature on the fact
that $AdS_5\times S^5$ superspace (the superspace with bosonic body
$AdS_5\times S^5$) is a coset of $SU(2,2|4)$ supergroup. Although
important, this does not change the fact that this '$AdS_5\times
S^5$ superstring' {\it is just a particular case of the GS
superstring in a curved superspace \cite{GSinCurv}}. So is its type
IIA counterpart, '$AdS_4\times \mathbb{CP}^3$ superstring', which
attracted recently much attention, but is not a model on a coset of
supergroup, just because the type IIA supergravity superspace with
the bosonic body $AdS_4\times \mathbb{CP}^3$ is not a coset
\cite{Dima+08}.

Thus we prefer to formulate our problem as superembedding
description of the GS superstring model in $AdS_5\times
S^5$ superspace. On one hand, this formalism can be applied to study
the {\cal N}=16 two dimensional supergravity induced on the
worldsheet superspace of the superstring moving in AdS superspace.
And, in this respect, it is proper for the present volume because
two dimensional gravity and supergravity model were always in the
center of Wolfgang's interests, see {\it
e.g.}\cite{Kummer:1995qh,Grumiller:2002nm,Bergamin:2003am}.

On the other hand, the results of this manuscript can be useful in
further study of classical and quantum $AdS_5\times S^5$
superstring, which is of current interest for the applications of
AdS/CFT correspondence \footnote{See, for instance,
\cite{Tseytlin08,fTdual} where the $AdS_5\times S^5$ superstring was
used to reveal the mysterious dual superconformal symmetry of the
{\cal N}=4 SYM amplitudes.}.

\section{Introduction}


The standard GS superstring action\cite{G+S84} is based on embedding of a bosonic surface
$W^{2}$ in the target superspace $\Sigma^{(D|n)}$ ($D=3,4,6,10$,
$n=2(D-2)$ for heterotic and type I and $n=4(D-2)$ for type II
superstrings). This embedding is described by the bosonic and
fermionic coordinate functions
\begin{eqnarray}
\label{W0inS} & W^{p+1}\in \Sigma^{(D|n)}\, : \quad
{\hat{Z}}^{{M}}(\xi)= (\hat{x}^{\mu}(\xi)\, , \,
\hat{\theta}^{\check{\alpha}}(\xi))\; , \qquad ^{ \mu=0,1,\ldots ,
(D-1),}_{\check{\alpha}=1,\ldots, n\; ,} \quad
\end{eqnarray}
where  p$=$1 and $\xi^m=(\tau, \sigma)$ are local coordinates on
$W^{2}$. The more `ancient' Ramond--Neveu--Schwarz (RNS) or spinning
string, which becomes equivalent to the GS sigma model
on the quantum level and after imposing the so--called GSO
projection (see, however, \cite{vz,stvz} and more recent
\cite{uvarov}), corresponds to an embedding of the worldsheet
superspace $W^{(2|1+1)}$ into the spacetime $M^{D}= \Sigma^{(D|0)}$,
described by D bosonic superfields $\hat{X}^{\underline{m}}(\xi ,
\eta , \bar{\eta} )= \hat{x}^{\underline{m}}(\xi) + i\eta
\psi^{\underline{m}}(\xi) + i \bar{\eta}
\bar{\psi}^{\underline{m}}(\xi)+ \ldots $ depending on two bosonic
($\xi^m$) and complex fermionic coordinate $\eta$ (or
real fermionic coordinate in the case of heterotic string). There
are some known obstacles for extending such a description to
supermembrane and other branes\cite{SpinSM,SpinBST}.

Following \cite{stv}, the {\it superembedding approach}, developed
in \cite{bpstv} for 10D superstrings and 11D supermembrane, and
applied in the first studies of dynamics of Dirichlet
$p$-branes (D$p$--branes) and M-theory 5-brane (M5-brane) in
seminal papers \cite{hs1} and \cite{hs2}, describes strings and
branes by {\it embedding of a worldvolume superspace $W^{(p+1|n/2)}$
into the target superspace $\Sigma^{(D|n)}$}.

Let us denote the $d=p+1\leq D$ local bosonic coordinates and $n/2$
fermionic coordinates of $W^{(p+1|n/2)}$ by $\zeta^{{\cal
M}}=(\xi^m,\eta^{\check{q}} )$. Then the embedding of
$W^{(p+1|n/2)}$ into the tangent superspace $\Sigma^{(D|n)}$ with
coordinates $Z^{{M}}= (x^{\mu} ,  \theta^{\check{\alpha}})$ can be
described parametrically by specifying the set of coordinate
super-functions, the {\it worldvolume superfields}
$\hat{Z}^{{M}}(\zeta)=\hat{Z}^{{M}}(\xi^m , \eta^{\check{q}})$
\begin{eqnarray}
\label{WinS} && W^{(p+1|n/2)}\in \Sigma^{(D|n)}\; : \quad Z^{{M}}=
\hat{Z}^{{M}}(\zeta)= (\hat{x}^{\mu}(\xi , \eta)\; ,  \;
\hat{\theta}^{\check{\alpha}}(\xi , \eta)\; )\; . \qquad
 \end{eqnarray}
 Here, $
 \mu= 0,1,\ldots , (D-1)$,  $\check{\alpha}=1,\ldots
 n$, $m=0,1,\ldots, p$ and $\check{q}=1,\ldots\, ,
{n\over 2}$. Notice that the number of fermionic `directions'
$\eta^{\check{q}}$ of the worldvolume superspace are usually chosen
to be one--half of the number of fermionic dimensions of the target
superspace .\footnote{\label{footnote-f} For ${\cal N}=1$
$n=\delta_{\check{\alpha}}^{\check{\alpha}}$ is the number of values
of the minimal D--dimensional spinor index; for $D\not= 2\;(mod\;
8)$ this is $n=2^{[D/2]}$, where is $[D/2]$ is the integer part of
$D/2$; and for $D= 2\;(mod\; 8)$ it is $n=2^{[D/2]-1}$.}
 This is proper to replace {\it all} the
$\kappa$--symmetries\cite{dA+L82,W.S.83,G+S84} of the {\it
standard}, Dirac--Nambu--Goto type super-p-brane actions{}
\cite{BST87}, {}\cite{AETW} by the local worldvolume supersymmetry
\footnote{Under the standard super-p-branes we mean supersymmetric
extended objects the ground state of which are 1/2 BPS states, {\it
i.e.} preserve 1/2 of the tangent space supersymmetry reflected by
n/2 parametric $\kappa$--symmetry of their worldvolume actions. See
\cite{BL98+} as well as \cite{Z+U02,B02+,BdAPV03} for the actions in
enlarged (tensorial) superspaces with additional tensorial
coordinates (see \cite{vH+VP82,BL98+,JdA00} and refs therein)
describing the excitations of $k/32$ BPS states, including the
$k=31$ models possessing the properties of BPS preons\cite{BPS01}.},
thus realizing the idea developed for D=3,4 superparticle in
\cite{stv} \footnote{See \cite{stvz}{}, \cite{Dima+2002} and refs.
therein for formulations of superbranes in the worldvolume
superspaces with less than $n/2$ fermionic 'directions'.}.

\subsection{Superembedding equation}

For all presently known superbranes the embedding (\ref{WinS}) of
their maximal worldvolume superspace $W^{(p+1|{n\over 2})}$ into the
target superspace $\Sigma^{(D|n)}$ obeys the {\it superembedding
equation}. For D=3,4 superparticle this was obtained in \cite{stv}
by varying a superfield action (called STV action in ninties). To
write its most general and universal form for a super--$p$--brane in
D-dimensional supergravity background, let us denote  the
supervielbein of the worldvolume superspace $W^{(p+1|16)}$ by
\begin{eqnarray}
\label{eA=ea+} e^A= d\zeta^{{\cal M}} e_{{\cal M}}{}^{A}(\zeta) =
(e^a\, , \, e^q) \, , \quad a=0,1,\ldots , p\; , \quad q=1,\ldots,
n/2 \; , \;
\end{eqnarray}
and decompose the pull--back
$\hat{E}^{\underline{A}}:=E^{\underline{A}}(\hat{Z})=d\hat{Z}^{{M}}
E_{{M}}{}^{{\underline{A}}}(\hat{Z})$ of the supervielbein of the
target superspace, $E^{\underline{A}}=dZ^{{M}}
E_{{M}}{}^{\underline{A}}(Z)=(E^{\underline{a}}, E^{{\alpha}})$
($\underline{a}=0,1,\ldots (D-1)$, $\alpha=1,\ldots , n$), on the
basis of (\ref{eA=ea+}). In general, such a decomposition reads
\begin{eqnarray}
\label{hEa=b+f}
 \hat{E}^{\underline{A}}:= E^{\underline{A}}(\hat{Z})=
d\hat{Z}^{{M}} E_{{M}}{}^{\underline{A}}(\hat{Z}) = e^b
\hat{E}_b^{\, \underline{A}} + e^q \hat{E}_q^{\, \underline{A}} \; ,
\qquad
\end{eqnarray}
where $\hat{E}_b^{\, \underline{A}}:= e_b^{{\cal M}} \partial_{{\cal
M}} \hat{Z}^{{M}} E_{{M}}{}^{\underline{A}}(\hat{Z})$ and
$\hat{E}_q^{\, \underline{A}}:= e_q^{{\cal M}}\partial_{{\cal M}}
\hat{Z}^{{M}} E_{{M}}{}^{\underline{A}}(\hat{Z})$ are, respectively,
bosonic and fermionic components of the pull--back the supervielbein
form. The superembedding equations states that the fermionic
component of the pull--back of the bosonic supervielbein form
vanishes,
\begin{eqnarray}
\label{SembEq}
 \fbox{$\hat{E}_q^{\,\underline{a}}:= \nabla_q
 \hat{Z}^{{M}}\,
E_{{M}}^{\;\,\underline{a}}(\hat{Z}) =0 \;$}\;  ,  \qquad
\nabla_q:=e_q^{{\cal M}}(\zeta) \partial_{{\cal M}}\; . \qquad \;
\end{eqnarray}

For higher dimensional superbranes of sufficiently large
co-dimensions the superembedding equation contains equations of
motion among their consequences. This was shown for  M2-brane and
D=10 type II superstring in \cite{bpstv}, for M5-brane in \cite{hs2}
and for D$p$--branes with $p\leq 5$ in \cite{hs1} (the 'boundary'
$p\leq 5$ was established in \cite{hs+}). Hence, in these cases, the
description of the classical super-$p$-brane dynamics by this
equation is complete. Moreover, if several types of $D$--dimensional
p-branes exist, the superembedding equation provides their universal
description (see \cite{IB2001} for such a universal description of
fundamental type IIB superstring and D1--brane and \cite{DpSL2} for
the $SL(2)$ covariant formulation  providing a unified descriptions
of all the actions of  p--branes related to the D$p$-brane by SL(2)
transformations).

On the other hand, this on-shell nature of the superembedding
equation prevents from the constructing the complete worldvolume
superfield action of the STV type (see \cite{stv} and \cite{Dima}
for the review and further references). A universal although
non-standard Lagrangian framework for the superembedding approach is
provided by the generalized action principle, proposed in \cite{bsv}
for superstrings and $D=11$ supermembrane and in \cite{bst} for the
case of super-D$p$-branes. This produces the superembedding equation
in its equivalent form
\begin{equation}\label{sEi=0}
\hat{E}^i(\zeta):= d\hat{Z}^{{M}}(\zeta)\,
{E}_{{M}}{}^{\underline{b}}(\hat{Z}(\zeta))\,
u_{\underline{b}}^{\;\; i} (\zeta) = 0 \; ,
\end{equation}
where $u_{\underline{b}}^{\;\; i}$ are $(D-p-1)$ vectors orthogonal
to the worldsheet superspace. These {\it moving frame variables} or
Lorentz harmonics (vector harmonics) will be the subject of the next
section.

\section{Spinor moving frame formulation and  generalized action principle for super-$p$-branes}
\label{sec1.2}

\subsection{Vector harmonics as moving frame  adapted to
(super)embedding} \label{mFrame}

The standard  formulations of superstring \cite{G+S84}, M2-brane
(supermembrane) \cite{BST87} and super-$p$-branes \cite{AETW} is
based on embedding (\ref{W0inS}) of the bosonic worldvolume
$W^{p+1}$ into the tangent superspace $\Sigma^{(D|n)}$.

If the worldvolume $W^{p+1}$ is flat, one always can chose a special
Lorentz frame with $p+1$ vectors being tangential and the remaining
$D-p-1$ vectors - orthogonal to $W^{p+1}$. In general this also can
be done, but locally. It is convenient to use the dual language of the differential forms and to consider  the pull--back
\begin{equation} \label{Eua:=def} \hat{E}^{\underline{a}}:= {E}^{\underline{a}}(\hat{Z})= d\hat{Z}^{{M}}(\xi)
E_{{M}}{}^{\underline{a}}(\hat{Z}) = d\xi^m
\partial_{m}\hat{Z}^{{M}} E_{{M}}{}^{\underline{a}}(\hat{Z}) =:
d\xi^m \hat{E}_m^{\;  \underline{a}}  \;
\end{equation}
of the bosonic supervielbein of the target superspace
${E}^{\underline{a}}:= d{Z}^{{M}} E_{{M}}{}^{\underline{a}}({Z})$ to
the worldvolume $W^{p+1}$ with local coordinates $\xi^m$,
$m=0,1,\ldots , p$. Only $(p+1)$ of the $D$ one--forms
$\hat{E}^{\underline{a}}$ may be independent on $W^{p+1}$. This is
tantamount to saying that there exist $(D-p-1)$ linear combinations
of $\hat{E}^{\underline{a}}$ that vanish on $W^{p+1}$. We can
express the above statement by  the following {\it embedding
equation}
\begin{equation}\label{Ei=0}
\hat{E}^i(\xi):= \hat{E}^{\underline{b}}\,  u_{\underline{b}}^{\;\;
i} (\xi) = 0 \; , \qquad i=1,\ldots , (D-p-1)\; , \quad
\end{equation}
where $u_{\underline{b}}^{\;\; i} (\xi)$ are some coefficient
dependent on the point of $W^{p+1}$. They define $(D-p-1)$ vectors
which are linear independent and orthogonal to the worldvolume
$W^{p+1}$. Thus one may chose them orthogonal one to another and
normalized   (on $-1$ as the vectors are spacelike and we are
working with 'mostly minus' metric conventions)
\begin{eqnarray}\label{uiuj=}
{u}^{\underline{a} i} u_{\underline{a}} ^{\; j} =-\delta^{ij}\; .
\end{eqnarray}

One can complete the set of the $(D-p-1)$ vectors $u_{\underline{a}}
^{\; i}$ orthogonal to the worldvolume by the set of the (p$+$1)
vectors $u_{\underline{a}} ^{\; b}$ tangential to $W^{p+1}$ (also
orthogonal among themselves and normalized). Then the $D\times D$
{\it moving frame matrix} constructed from $ u_{\underline{a}} ^{\;
b}$ and $ u_{\underline{a}} ^{\; j}$ obeys $U\eta U^{\!^T}=\eta$,
\begin{eqnarray}\label{UTIU=I}
U_{\underline{a}}^{(\underline{b})} := \left(\begin{matrix}
u_{\underline{a}} ^{\; b}\cr u_{\underline{a}} ^{\; j}\end{matrix}
\right) \; , \qquad  U^{\!^T} \eta U=\eta \quad \Leftrightarrow
\qquad  \begin{cases} {u}^{\underline{c} a} u_{\underline{c}} ^{\;
b} =\eta^{ab}\; , \cr {u}^{\underline{a} a} u_{\underline{a}} ^{\;
j} =0\; , \cr {u}^{\underline{a} i} u_{\underline{a}} ^{\; j}
=-\delta^{ij}\; , \end{cases}
\end{eqnarray}
by construction, and, hence, belongs to the fundamental
representation of the Lorentz group $SO(1,D-1)$,
\begin{eqnarray}\label{uIu=I}
U_{\underline{a}}^{(\underline{b})} := \left(\begin{matrix}
u_{\underline{c}} ^{\; b}\cr u_{\underline{c}} ^{\; j}\end{matrix}
\right) \; \in  \; SO(1,D-1) \; .
\end{eqnarray}

The splitting of the $D\times D$ matrix $U$ on the  D$\times$(p+1)
and D$\times$(D$-$p$-$1) blocks (\ref{uIu=I}) is invariant under the
(right multiplication by the matrix from the) $SO$(1,p)$\otimes
SO$(D$-$p$-$1) subgroup of the Lorentz group $SO(1,$D$-$1). In the
Lorentz harmonic approach of \cite{BZ-str,BZ-p} \footnote{See
\cite{Sok}, \cite{niss}, \cite{K+R88},  {} \cite{Wiegmann}, {}
\cite{B90}, {} \cite{gds} and \cite{ghs} for earlier works.} this
gauge invariance is usually considered as an identification relation
on the set of moving frame variables making possible to consider
them as  'homogeneous' coordinate for the coset
\begin{eqnarray}\label{u-in-coset}
 \left\{u_{\underline{a}} ^{\; b} \, , \,
u_{\underline{a}} ^{\; j} \right\} \quad  =  \qquad
{SO(1,D-1)\over SO(1,p) \otimes SO(D-p-1) }\; .
\end{eqnarray}
This was the reason to call these moving frame variables {\it
Lorentz harmonics}  \cite{B90,BZ-str}, following the spirit of
\cite{GIKOS} where the notion of harmonic variables was introduced
to construct the unconstrained superfield formulation of the ${\cal
N}=2$ supersymmetric theories.

Reordering the line of arguments one can start from
(\ref{u-in-coset}) and notice that $SO$(1,D$-$1) group valued moving
frame matrix $U$ (\ref{uIu=I}) can be used to define, starting from
$\hat{E}^{(\underline{a})}$, another vielbein attached to the
worldvolume,
\begin{eqnarray}\label{Eua=Ea,Ei}
\hat{E}^{(\underline{a})} :=
\hat{E}^{\underline{b}}U_{\underline{b}}^{(\underline{a})}=:
(\hat{E}^{a} , \hat{E}^{i}) \; .
\end{eqnarray}
This vielbein is adapted to the embedding of $W^{p+1}$ into the
$D$--dimensional spacetime if the pull--back of $D-p-1$ `orthogonal'
forms
\begin{eqnarray}\label{Ei=Eui}
\hat{E}^{i} := \hat{E}^{\underline{a}}u_{\underline{a}}^i \;
\end{eqnarray}
vanishes, {\it i.e.} if embedding equation (\ref{Ei=0}) is valid.
The $(p+1)$ 'tangential' forms $\hat{E}^a$ defined with the use of
the 'parallel' vector harmonics $u_{\underline{b}} ^{\; a}$
\begin{eqnarray}\label{Ea=Eua}
\hat{E}^{a} := \hat{E}^{\underline{b}}u_{\underline{b}}^a \;
\end{eqnarray}
can be used as a vielbein on $W^{p+1}$;
\begin{eqnarray}\label{ea=Eua}
e^a = \hat{E}^{a} := \hat{E}^{\underline{b}}u_{\underline{b}}^a \; .
\end{eqnarray}
One says that this vielbein is induced by the embedding.

Now one sees that the superembedding equation (\ref{sEi=0}) is just
the straightforward supersymmetric generalization of the above
embedding equation (\ref{Ei=0}). However, in contrast to
(\ref{Ei=0}), the superembedding equation  cannot be derived by
imposing a conventional orientation conditions, and in this sense is
nontrivial.

\subsection{Action of the  moving frame formulation}

This is the induced vielbein (\ref{ea=Eua}) which can be understood
as a square root from the induced metric
\begin{eqnarray}\label{gmn=EmEn}
g_{mn}(\xi)= \hat{E}_m^{\underline{a}} \hat{E}_{n\,\underline{a}}
\end{eqnarray}
provided the embedding equation (\ref{Ei=0}) holds,
\begin{eqnarray}\label{g=EaEa}
E^i=0 \quad \Rightarrow \quad g_{mn}(\xi)= \hat{E}_m^{\underline{a}}
\hat{E}_{n\,\underline{a}}= \hat{E}_m^{{a}} \hat{E}_{n\,{a}}=e_m^{\;
{a}} e_{n\,{a}}\;  .
\end{eqnarray}
As a result,  the invariant volume element on $W^{p+1}$, this is to
say the Nambu--Goto term for a (super)--$p$--brane,
\begin{eqnarray}\label{SNG=}
S^{N-G}_p:= \int d^{p+1} \xi \sqrt{|g|}:= \int d^{p+1} \xi
\sqrt{|det(\hat{E}_m^{\underline{a}} \hat{E}_{n\,\underline{a}}) |}
 \; .
\end{eqnarray}
can be equivalently presented in terms of $e^a:= \hat{E}^a$ forms,
$\int_{W^{p+1}} \hat{E}^{\wedge ({p+1})}$,
\begin{eqnarray}\label{g=EvE}
E^i=0 \quad \Rightarrow  \quad d^{p+1} \xi \sqrt{|g|}:=
{\varepsilon_{a_0 \ldots a_p } \hat{E}^{{a}_0}\wedge \ldots \wedge
\hat{E}^{{a}_p}\over (p+1)!} =: \hat{E}^{\wedge ({p+1})}
 \; . \quad
\end{eqnarray}
Now, if one uses $\hat{E}^{\wedge ({p+1})}$ instead of the
Nambu--Goto term (\ref{g=EvE}) in the standard super--$p$--brane action
 {}\cite{AETW},
\begin{eqnarray}\label{Sp-br-st}
 S^{standard}_p= S^{N-G}_p+S_p^{WZ}&:=& \int d^{p+1} \xi \sqrt{|g|} - p  \int\limits_{W^{p+1}} \hat{B}_{p+1} \, , \qquad
\end{eqnarray}
one arrives at the so--called {\it
moving frame} or {\it Lorentz harmonic} action
\begin{eqnarray}\label{Sp-br-LH}
  S_p &=& S^{LH}_p+S_p^{WZ}:= \int\limits_{W^{p+1}} \hat{E}^{\wedge ({p+1})} -
  p \int\limits_{W^{p+1}} \hat{B}_{p+1} \,  \qquad
 \;  \qquad \nonumber
 \\  &=& \int\limits_{W^{p+1}}  {1\over (p+1)!} \varepsilon_{a_0 \ldots a_p
}\hat{E}^{{a}_0}\wedge \ldots \wedge \hat{E}^{{a}_p} - p
 \int\limits_{W^{p+1}} \hat{B}_{p+1} \; ,   \qquad
  \end{eqnarray}
where $\hat{E}^{a} = \hat{E}^{\underline{b}}u_{\underline{b}}^a$,
Eq. (\ref{Ea=Eua}) and $u_{\underline{b}}^a$ are $(p+1)$ orthonormal
$D$--vectors, $u^{\underline{b}a}u_{\underline{b}}^b=\eta^{ab}$ (see
 (\ref{UTIU=I})). These are the auxiliary
variable entering the action without derivatives. The last term of
the standard action, $-p \int\limits \hat{B}_{p+1} $, which remains
in the same form in the spinor moving frame formulation, is the
so--called Wess--Zumino (WZ) term. It is given by the integral of
the pull--back the worldvolume $W^{p+1}$ of the gauge
$(p+1)$--superform ${B}_{p+1}$ restricted by the superspace
constraints imposed on its (super)field strength
\begin{eqnarray} \label{Hp+2}
& H_{p+2} = dB_{p+1}= \propto \bar{\Gamma}^{(p)}_{\alpha\beta}
\wedge E^{\alpha} \wedge E^{\beta} + {\cal O} (E^{\wedge (p+1)}) \;
, \qquad \label{bGp} \\ & \bar{\Gamma}^{(p)}_{\alpha\beta}:= {1
\over p!}
 E^{\underline{a}_1} \wedge \ldots \wedge
E^{\underline{a}_{p}}\,  \Gamma_{\underline{a}_p \ldots
\underline{a}_{1}\; \alpha\beta } \; .
\end{eqnarray}
The relation between coefficient for the first term in the {\it
r.h.s.} of (\ref{Hp+2}) (replaced by $\propto$ symbol in our
schematic consideration) and the coefficient in front of the WZ term
in the action is fixed by the requirement of $\kappa$--symmetry.

As it has been noticed above, on the surface of embedding equation
(\ref{Ei=0}) the moving frame action (\ref{Sp-br-LH}) coincides with
the standard one, Eq.(\ref{Sp-br-st}),
\begin{eqnarray}\label{S(Ei=0)=S}
&& S_p\vert_{{{\hat{E}^i}=0}}= S_p^{standard} \; .
\end{eqnarray}
The proof of the classical equivalence will then be completed by
showing that the embedding equation follows from the moving frame
action (\ref{Sp-br-LH}).

This is indeed the case, the embedding equation appears as a result
of varying the auxiliary moving frame variables in the action (\ref{Sp-br-LH}),
\begin{eqnarray}\label{du-S-Ei}
\delta_{_{\delta u}} S_p=0 \qquad \Rightarrow \qquad {\hat{E}^i}:=
\hat{E}^{\underline{a}} u_{\underline{a}}{}^i =0 \; .
\end{eqnarray}
As the harmonics are constrained variables, the variation in  Eq.
(\ref{du-S-Ei}) requires some comments.

\subsubsection{Variations and derivatives of the harmonic variables}
\label{vectorH}

Both the spaces of the variations $\delta u$ of certain variables
$u$ and of the derivatives $du$ of such variables can be identified
with the elements of the fiber of the tangent bundle over the space
of this variables, i.e. with elements of the linear space tangent to
the space of the $u$ variables. In the case of Lorentz harmonics the
variables $u$ are elements of the Lorentz group valued matrix $U$,
Eq. (\ref{uIu=I}) (see also (\ref{UTIU=I})). The space tangent to
the Lorentz group is isomorphic to the Lie algebra spanned by
antisymmetric $D\times D$ matrices. This well known fact can be
expressed by
\begin{eqnarray}\label{dU=UOm}
dU_{\underline{a}}^{(\underline{b})}=
U_{\underline{a}}^{(\underline{c})}
\Omega_{(\underline{c})}{}^{(\underline{d})}\equiv
U_{\underline{a}\, (\underline{c})}
\Omega^{(\underline{c})(\underline{b})}
 \quad
\Leftrightarrow \qquad  \begin{cases} du_{\underline{a}} ^{\; b} =
u_{\underline{a}c} \Omega^{c b} + u_{\underline{a}}^{\; i}
\Omega^{bi }\; ,
 \cr  du_{\underline{a}} ^{\; i}= - u_{\underline{a}}^{\; j}\Omega^{ji} +
 u_{\underline{a}b} \Omega^{bi }\end{cases}\; ,
\end{eqnarray}
which is just an equivalent representation of the definition of the
Cartan forms $U^{-1}dU$ for the Lorentz group in which $U^{-1}=\eta
U^T \eta $,
\begin{eqnarray}\label{UdU=Om}
U^{\underline{c}(\underline{a})}
dU_{\underline{c}}^{(\underline{b})} &=:&
\Omega^{(\underline{a})(\underline{b})}\equiv -
\Omega^{(\underline{b})(\underline{a})} = \left\{
\begin{matrix}\; \;\Omega^{ab} & \Omega^{aj} \cr -\Omega^{bi} & \Omega^{ij}  \end{matrix} \right\}\,
  . \quad
\end{eqnarray}
As far as the harmonics are treated as homogeneous  coordinates of
the coset ${SO(1,D-1) \over SO(1,p)\otimes SO(D-p-1)}$, the Cartan
forms $\Omega^{ab}=-\Omega^{ba}=u^{\underline{c}\,
a}du_{\underline{c}} ^{\; b}$ and
$\Omega^{ij}=-\Omega^{ji}:=u^{\underline{c}\, i}du_{\underline{c}}
^{\; j}$ have the properties of the connection under the $SO(1,p)$
and $SO$(D$-$p$-$1) local gauge symmetries  while the set of
one--forms $\Omega^{ai}=u^{\underline{c}\, a}du_{\underline{c}} ^{\;
i}$ provide the vielbein for the coset ${SO(1,D-1) \over
SO(1,p)\otimes
  SO(D-p-1)}$.

  One should notice that, in general,  the transformation of
$SO(1,p)$ and $SO$(D$-$p$-$1) symmetries are local on the coset
space itself; however, when harmonics are used to describe a
$p$--brane, $U=U(\xi)$, this local $SO(1,p) \otimes SO$(D$-$p$-$1)
symmetry of Lorentz harmonic space (or superspace) gives rise to the
worldvolume local gauge symmetries with $\xi$--dependent parameters.

The same line of reasoning can be applied to the variations of the
Lorentz harmonic variables in some action functional. Formally, the
corresponding equation can be derived by using the Lie derivative
${\cal L}_\delta := i_\delta d + d i_\delta$ where the second terms
will give zero contributions for a zero--forms so that $\delta u=
i_\delta du$. Applying this simple equation to (\ref{dU=UOm}) one
finds
\begin{eqnarray}\label{vU=UOm}
\delta U_{\underline{a}}^{(\underline{b})}=
U_{\underline{a}}^{(\underline{c})} i_\delta
\Omega_{(\underline{c})}{}^{(\underline{b})}
 \quad
\Leftrightarrow \qquad  \begin{cases}\delta u_{\underline{a}} ^{\;
b} = u_{\underline{a}c} i_\delta \Omega^{c b} +
u_{\underline{a}}^{\; i} i_\delta \Omega^{bi }\; ,
 \cr  \delta u_{\underline{a}} ^{\; i}= - u_{\underline{a}}^{\; j}i_\delta \Omega^{ji} +
 u_{\underline{a}b} i_\delta \Omega^{bi }\end{cases}\; ,
\end{eqnarray}
where $i_\delta \Omega^{(\underline{a})(\underline{b})}= - i_\delta
\Omega^{(\underline{b})(\underline{a})} = \{ i_\delta \Omega^{a b}\,
, i_\delta \Omega^{ij } , i_\delta \Omega^{aj } \}$ are parameters
of independent variations which can be identified with the
$i_\delta$ contractions of the Cartan forms (hence the notation).
Clearly $i_\delta \Omega^{a b}= - i_\delta \Omega^{ba}$ parametrize
the worldsheet Lorenz group $SO(1,p)$, $i_\delta \Omega^{a b}= -
i_\delta \Omega^{ba}$ parametrize the transformations  of the
structure group $SO$(D$-$p$-$1)  of the normal bundle and $i_\delta
\Omega^{bi}$ provides a basis for independent variations of the
coset ${SO(1,D-1) \over SO(1,p)\otimes
  SO(D-p-1)}$.

\subsubsection{Lorentz harmonics and generalized Cartan forms for
superbrane in curved (super)space} \label{CartanF}

In the curved superspace one has to consider the local Lorentz
$SO$(1,D$-$1) symmetry and the Cartan forms as defined in
(\ref{UdU=Om}) or (\ref{dU=UOm}) are not covariant. Their covariant
counterparts are defined with the use of  Lorentz covariant
derivatives $(d+w)$, so that
\begin{eqnarray}\label{DU=UOm}
  \begin{cases} Du_{\underline{a}} ^{\; b} := du_{\underline{a}} ^{\; b} +
u^{\underline{c}\, i} \hat{w}_{\underline{c}}{}^{\underline{a}} -
u_{\underline{a}}{}^c \Omega_c{}^{b} = u_{\underline{a}}^{\; i}
\Omega^{bi }\; ,
 \cr  Du_{\underline{a}} ^{\; i} := du_{\underline{a}} ^{\; b} +
u^{\underline{b}\, i} \hat{w}_{\underline{b}}{}^{\underline{a}} +
u_{\underline{a}}^{\; j}\Omega^{ji} =
 u_{\underline{a}b} \Omega^{bi } \; , \end{cases}\qquad
\end{eqnarray}
or
\begin{eqnarray}\label{UDU=Om}
\Omega^{(\underline{a})(\underline{b})} &:=&
U^{\underline{c}(\underline{a})}
[(d+\hat{w})U]_{\underline{c}}^{(\underline{b})}
 \quad
 \Leftrightarrow \quad  \nonumber \\ && \begin{cases} \Omega^{a b}= u^{\underline{c}\, a}[(d+\hat{w}) u]_{\underline{c}} ^{\; b}
 = \Omega^{0\, a b} + (u\hat{w}u)^{ab} \; , \quad (a)
 \cr \Omega^{ij}:= u^{\underline{c}\, i}[(d+\hat{w}) u]_{\underline{c}} ^{\; j}= \, \Omega^{0\,
 ij} +  (u\hat{w}u)^{ij} \; , \quad  (b)
 \cr \Omega^{ai } :=  u^{\underline{c}\, a}[(d+ \hat{w})u]_{\underline{c}} ^{\; i}=
  \Omega^{0\, ai } +  (u\hat{w}u)^{ai} \; , \quad (c)\end{cases}
  , \quad
\end{eqnarray}
where
$\hat{w}=d\hat{Z}^{{M}}w_{{M}\,\underline{b}}{}^{\underline{a}}(\hat{Z})$
is the pull--back of the tangent superspace spin connection
$w_{\underline{b}}{}^{\underline{a}}=
d{Z}^{{M}}w_{{M}\,\underline{b}}{}^{\underline{a}}({Z})$
 and $\Omega^{0\; ai }$,
$\Omega^{0\; ij }$, $\Omega^{0\; ai }$ are the Cartan forms as
defined in  Eqs. (\ref{UdU=Om}). Then $D$ in (\ref{DU=UOm}) is the
derivative covariant both with respect to the local Lorentz
$SO(1,D-1)$ transformations and worldsheet gauge symmetry
$SO(1,p)\otimes SO(D-p-1)$. This implies that
\begin{eqnarray}\label{DEa=}
D\hat{E}^a := d\hat{E}^a - \hat{E}^b \wedge \Omega_b{}^a =
\hat{T}^{\underline{b}} u_{\underline{b}}{}^a + \hat{E}^i \wedge
\Omega^{ai} \; ,
  \quad
\end{eqnarray}
where $\Omega^{ai}$ is the generalized Cartan form  (\ref{UDU=Om}c)
and $\hat{T}^{\underline{b}}$ is the pull--back of the superspace
torsion ${T}^{\underline{a}} =
D{E}^{\underline{a}}=d{E}^{\underline{a}}- {E}^{\underline{b}}\wedge
w_{\underline{b}}{}^{\underline{a}}$. This is restricted by
supergravity constraints
\begin{eqnarray}\label{Ta=constr}
{T}^{\underline{a}} = D{E}^{\underline{a}} := d{E}^{\underline{a}} -
{E}^{\underline{b}}\wedge w_{\underline{b}}{}^{\underline{a}} =  -i
E^{\alpha} \wedge E^{\beta} \Gamma^{\underline{a}}_{\alpha\beta} \;
\end{eqnarray}
 With this in mind one can further specify Eq. (\ref{DEa=})
\begin{eqnarray}\label{DEa=EsE}
D\hat{E}^a  = -i \hat{E}^{\alpha} \wedge \hat{E}^{\beta}
\hat{\Gamma}^a_{\alpha\beta} \;
 + \hat{E}^i \wedge \Omega^{ai} \; , \qquad
\hat{\Gamma}^a:= \Gamma^{\underline{b}}_{\alpha\beta} \;
 u_{\underline{b}}{}^a\; .
  \quad
\end{eqnarray}

Using the formal $i_\delta$ symbol of the previous subsection
[extending its definition by $i_\delta d\hat{Z}^{\cal M}:= \delta
\hat{Z}^{\cal M}$ and $i_\delta (\Omega_q \wedge \Omega_p) =
\Omega_q \wedge i_\delta\Omega_p + (-)^p i_\delta \Omega_q \wedge
\Omega_p$ for any q- and p-forms $\Omega_q$ and $\Omega_p$] one can
write the arbitrary variation of the 'tangential' supervielbein
$\hat{E}^a$ (modulo the $SO(1,p)$ symmetry transformations) in the
form
\begin{eqnarray}\label{varEa=}
\delta \hat{E}^a = D(i_\delta \hat{E}^a)\,  - 2i \hat{E}^{\alpha}\,
\hat{\Gamma}^a_{\alpha\beta} \, i_\delta \hat{E}^{\beta}
 + \hat{E}^i \, i_\delta \Omega^{ai} -  \Omega^{ai}\,  i_\delta \hat{E}^i \; , \qquad
\end{eqnarray}
where $i_\delta \Omega^{ai}$ are basic variation of the harmonic
variables, Eq. (\ref{vU=UOm}), corresponding to the coset
${SO(1,D-1) \over SO(1,p)\otimes SO(D-p-1)}$ and
\begin{eqnarray}\label{idEa:=}
&& i_\delta \hat{E}^a = \delta \hat{Z}^M\,  E_M{}^{\underline{c}}
(\hat{Z})\,
   u_{\underline{c}}{}^a(\xi)  \; ,
\qquad   i_\delta \hat{E}^i = \delta \hat{Z}^M\, E_M^{\;
\underline{a}} (\hat{Z})\,
   u_{\underline{a}}{}^i(\xi)  \; , \qquad \\
 \label{idEf:=} && \qquad i_\delta \hat{E}^{\alpha} =\delta \hat{Z}^M\,  E_M^{\; {\alpha}}
(\hat{Z})\, .  \qquad
\end{eqnarray}
These provide the covariant basis for the variations of the bosonic
and fermionic  coordinate functions $\delta \hat{Z}^M$.

\subsubsection{Equations of motion of the moving frame action}
\label{Eqm}

In the above notation general  variation of the action with respect
to the coordinate functions and harmonic variables,
\begin{eqnarray}\label{Sp=}
\delta S_p &=& \int\limits_{W^{p+1}} [\hat{E}^{ \wedge p}_a \wedge
\delta \hat{E}^a - p\delta\hat{B}_{p+1}] \;  \qquad
\end{eqnarray}
(with $\hat{E}^{ \wedge p}_a:= {1\over p!}\varepsilon_{aa_1\ldots
a_p} \hat{E}^{{a_1}} \wedge \ldots \wedge \hat{E}^{{a_p}}$), can be
written as
\begin{eqnarray}\label{vSp=}
\delta S_p =& \int\limits_{W^{p+1}} \hat{E}^{ \wedge p}_a \wedge
 [D i_\delta \hat{E}^a + i_\delta \hat{T}^a + \hat{E}^i \, i_\delta \Omega^{ai} -
 \Omega^{ai}\,  i_\delta \hat{E}^i ] \,
- p \int\limits_{W^{p+1}} i_\delta\hat{H}_{p+2} \; .  \qquad
\end{eqnarray}
Using (\ref{Ta=constr}), (\ref{Hp+2}) and  the identity
$\hat{E}^{^{a_1}}\!\! \wedge\! \ldots \!\wedge\! \hat{E}^{^{a_p}}
\hat{\Gamma}_{a_1\ldots a_p} \hat{\Gamma}_a = \propto
\hat{E}^{\wedge p}_a \,\bar{\hat{\Gamma}}$ with
\begin{eqnarray}\label{bGamma}
& \bar{\hat{\Gamma}} := i^{^{p(p-1)\over 2}}\hat{\Gamma}_0 \ldots
\hat{\Gamma}_p \equiv {i^{^{p(p-1)\over 2}} \over (p+1)!}\;
\varepsilon_{a_0 \ldots a_p} u_{\underline{b}_0}^{\; a_0} \ldots
u_{\underline{b}_p}^{\; a_p} \Gamma^{\underline{b}_0 \ldots
\underline{b}_p}  \;  \qquad
\end{eqnarray}
obeying $\bar{\hat{\Gamma}}\bar{\hat{\Gamma}} =I $, $tr
\bar{\hat{\Gamma}}=0$, one finds\footnote{We do not write explicitly
the terms proportional to the `tangential' bosonic variations
$i_\delta \hat{E}^{{a}}$, denoting them in (\ref{vfSp}) by ${\cal
O}(i_\delta \hat{E}^{{a}})$, as they do not produce any independent
equation. This statement manifests a Noether identity which
corresponds to the reparametrization gauge symmetry, {\it i.e.} the
worldvolume diffeomorphism invariance.}
\begin{eqnarray}\label{vfSp}
\delta S_p &= -2i \int\limits_{W^{p+1}}
 \hat{E}^{ \wedge p}_a \wedge \hat{E}^\beta [\hat{\Gamma}^a
(I-\bar{\hat{\Gamma}})]_{\beta\alpha} \, i_\delta
 E^{\alpha}(\hat{Z}) +  \int\limits_{W^{p+1}} \hat{E}^{ \wedge p}_a \wedge \hat{E}^i \,
 i_\delta \Omega^{ai} - \qquad \nonumber \\ &- \int\limits_{W^{p+1}}
 (\hat{E}^{ \wedge p}_a \wedge \Omega^{ai} +
 p\, u^{{\underline{a}}i}\, i_{\underline{a}}\hat{H}_{p+2} )
 i_\delta \hat{E}^i \,
 + {\cal O}(i_\delta \hat{E}^{{a}}) \;  \qquad
\end{eqnarray}
The second term in (\ref{vfSp}) contains the basic variations of the
harmonic variables and is used to obtain the embedding equation
(\ref{Ei=0}). The third term produces the bosonic equations of
motion of the $p$--brane in the form
\begin{eqnarray}\label{Omaai=}
\Omega_a{}^{ai}= \propto \varepsilon^{a_1\ldots a_{p+1}}
H_{a_1\ldots a_{p+1}}{}^i+ fermion\; contributions \; ,
\end{eqnarray}
which generalizes the minimal surface equation $\Omega_a{}^{ai}=0$
for the case of nonvanishing background flux (see sec.
\ref{sec1.3.4} for more details in p$=$1 case). Finally the first
term in (\ref{vfSp})  contains the fermionic variation $i_\delta
\hat{E}^\alpha:= \delta \hat{Z}^{{M}} {E}_{{M}}{}^\alpha(\hat{Z})$
and produces the fermionic equation for super-$p$--brane
\begin{eqnarray}\label{fEq-p}
\Psi_{p+1\; \alpha} (\hat{Z}) :=  \hat{E}^{ \wedge p}_a \wedge
\hat{E}^\beta [\hat{\Gamma}^a (I-\bar{\hat{\Gamma}})]_{\beta\alpha}
=0 \; . \qquad
\end{eqnarray}

\subsection{Irreducible $\kappa$--symmetry.
 Spinor harmonics enter the game}
\label{KapIrr}

The presence of the  projector $(I-\bar{\hat{\Gamma}})$ makes half
of the fermionic equations (\ref{fEq-p}) to be satisfied
identically,
\begin{eqnarray}\label{k-NI}
\Psi_{p+1\;} (\hat{Z}) (I+\bar{\hat{\Gamma}}) \equiv 0 \; . \qquad
\end{eqnarray}
Eq. (\ref{k-NI}) is the Noether identity reflecting a fermionic
gauge symmetry of the action (\ref{Sp-br-LH}), the
$\kappa$--symmetry with the basic variations
\begin{eqnarray}\label{k-sym}
&& i_\kappa \hat{E}^{\underline{\; a}}:= \delta_\kappa
\hat{Z}^{{M}}(\xi) {E}_{{M}}{}^{\underline{\; a}}(\hat{Z}(\xi))= 0
\; , \qquad \nonumber
\\ && i_\kappa \hat{E}^\alpha:= \delta_\kappa
\hat{Z}^{{M}}(\xi) {E}_{{M}}{}^\alpha(\hat{Z}(\xi))=
(I+\bar{\hat{\Gamma}})^\alpha{}_\beta \kappa^\beta(\xi)\; .  \qquad
\end{eqnarray}
These are {\it formally} the same as the ones for the infinitely
reducible $\kappa$--symmetry of the standard action
(\ref{Sp-br-st}).  However, the presence of additional variables
makes the $\kappa$--symmetry of the action (\ref{Sp-br-LH}) {\it
irreducible} in contradistinction to the $\kappa$--symmetry of the
original action (\ref{Sp-br-st}).

To see this one should notice that, allowing for additional
variables,  one can {\it factorize} the $\kappa$--symmetry
projector. Within the Lorenz harmonic approach such a factorization
reads
\begin{eqnarray}\label{G-fact(V)}
(I
{}^{_{\;_{-}}}_{^{^{(+)}}}\bar{\hat{\Gamma}})^{{\alpha}}{}_{{\beta}}
= 2v_{{\beta}}{}^{\dot{\bar{\alpha}}\dot{q}} \;
v_{\dot{\bar{\alpha}}\dot{q}}{}^{{\alpha}} \; , \qquad (I
{}^{_{\;_{+}}}_{^{^{(-)}}}\bar{\hat{\Gamma}})^{{\alpha}}{}_{{\beta}}
= 2v^{{\bar{\alpha}}{q}}_{{\beta}} \;
v_{\bar{\alpha}{q}}{}^{{\alpha}}\; ,
\end{eqnarray}
where $\bar{\alpha}$, $\dot{\bar{\alpha}}$ are indices of the spinor
representations of $SO(1,p)$ (the same or different depending on the
values of $D$ and $p$) and $q$, $\dot{q}$ are indices of the (same
or different) representations of $SO(D-p-1)$ and
\begin{eqnarray}\label{VinSpin}
V_{{\beta}}^{({\alpha})} := \left( v^{{\bar{\alpha}}{q}}_{{\beta}}
\, , \, v^{\dot{\bar{\alpha}}\dot{q}}_{{\beta}} \right)\quad \in
\quad Spin(1, D-1) \qquad
\end{eqnarray}
is the $Spin(1,D-1)$--valued matrix of the {\it spinor moving frame
variables} or {\it spinor harmonics}. These variables are, the
`square root' of the vector harmonics (\ref{uIu=I}), (\ref{UTIU=I})
in the sense of that the following constraints hold
\begin{eqnarray}\label{VGVT=uHa}
V\Gamma^{(\underline{a})}V^T = \Gamma^{\underline{b}}
U_{\underline{b}}{}^{(\underline{a})} \qquad \Rightarrow \qquad
\begin{cases}V\Gamma^{a}V^T = \Gamma^{\underline{b}}
u_{\underline{b}}{}^{a}\; , \cr V\Gamma^{i}V^T =
\Gamma^{\underline{b}} u_{\underline{b}}{}^{i}\end{cases}
 \; .  \qquad
\end{eqnarray}
Eqs. (\ref{VGVT=uHa}) express the well known fact of that the
gamma--matrices are Lorentz invariant. An equivalent form of these
constraints is given by
\begin{eqnarray}\label{VTGV=uHa}
(V^{-1})^T\tilde{\Gamma}^{\underline{a}}V^{-1} =
\Gamma^{(\underline{b})} U_{(\underline{b})}{}^{\underline{a}}=
 \Gamma^{b} u_{b}{}^{\underline{a}} - \Gamma^{i} u^{i\underline{a}}\; ,  \qquad
\end{eqnarray}
where $V^{-1}:= V_{(\underline{\alpha})}{}^{\!\!\underline{\beta}}$
is the matrix inverse to
 (\ref{VinSpin}),
\begin{eqnarray}\label{V-1:=}
V_{({\alpha})}{}^{\!\!{\beta}} := \left(
v_{{\bar{\alpha}}\,{q}}{}^{\!\beta} \, , \,
v_{\dot{\bar{\alpha}}}{}_{\dot{q}}^\beta \right)\quad : \quad
V_{(\alpha)}^{\;\;\; \gamma} V_{\gamma}^{(\beta)}
=\delta_{(\alpha)}^{\;\; (\beta)}:=
\left(\begin{matrix}\delta_{{\bar{\alpha}}}{}^{{\bar{\beta}}}
\delta_{{q}}^{{p}} & 0\cr 0 &
\delta_{\dot{\bar{\alpha}}}{}^{\dot{\bar{\beta}}}
\delta_{\dot{q}}^{\dot{p}} \end{matrix}\right)\; .
\end{eqnarray}

The spinor moving frame variables are also called {\it spinorial
harmonics} because they provide the homogeneous coordinates for the
coset of $Spin(1,D-1)$ group doubly covering the coset of Eq.
(\ref{u-in-coset}),
\begin{eqnarray}\label{V=(v-v)}
\{V_{{\beta}}^{({\alpha})}\} := \left\{\left(
v^{{\bar{\alpha}}{q}}_{{\beta}} \, , \,
v^{\dot{\bar{\alpha}}\dot{q}}_{{\beta}} \right)\right\} =
{Spin(1,D-1)\over Spin(1,p)\otimes Spin(D-p-1)}\; .
\end{eqnarray}

Now, using the factorization (\ref{G-fact(V)}) one can write the
$\kappa$--symmetry transformations (\ref{k-sym}) in the {\it
irreducible} form
\begin{eqnarray}\label{k-IRR}  i_\kappa
\hat{E}^\alpha:= \delta_\kappa \hat{Z}^{{M}} {E}_{{M}}^{\;
\alpha}(\hat{Z})=  2 v_{{\bar{\alpha}}{q}}{}^{\!\alpha}\,
\kappa^{{\bar{\alpha}}{q}}
 \; , \qquad
\end{eqnarray}
where the {\it irreducible $\kappa$--symmetry parameter} is
\begin{eqnarray}\label{k-IRRp}
\kappa^{{\bar{\alpha}}{q}} :=
v^{{\bar{\alpha}}{q}}_\beta\kappa^\beta(\xi)\; .
\end{eqnarray}

\subsection{Spinor moving frame formulation of super-$p$-branes}

The spinor moving frame formulation of super-$p$-brane is described
by the moving frame action (\ref{Sp-br-LH}),
\begin{eqnarray}\label{Sp=sp-mfr}
  S_p = & \int\limits_{W^{p+1}} \hat{E}^{\wedge ({p+1})} -
  p \int\limits_{W^{p+1}} \hat{B}_{p+1} \; ,   \qquad
\end{eqnarray}
in which the vector Lorentz harmonics  $u_{\underline{b}}^a$ (moving
frame variables), entering the definition of $E^a$ in
$\hat{E}^{\wedge ({p+1})}:= {1\over (p+1)!} \varepsilon_{a_0 \ldots
a_p }\hat{E}^{{a}_0}\wedge \ldots \wedge \hat{E}^{{a}_p} $, are
composites of spinor harmonics as defined by the gamma--trace parts
of the constraints (\ref{VGVT=uHa}),
 \begin{eqnarray}\label{Ea:=EuaP}
\hat{E}^{a} := \hat{E}^{\underline{b}}u_{\underline{b}}^a \; ,
\qquad & u_{\underline{b}}{}^a = {1\over n} tr
(\tilde{\Gamma}_{\underline{b}}V\Gamma^aV^T)\; ,
\end{eqnarray}
where  $n$ the number of values of (minimal) D--dimensional spinor
indices, $n=\delta_{{\alpha}}{}^{\alpha}$ (see footnote
\ref{footnote-f}). This composite nature does not change the
variation of vector harmonics, which are expressed through the
Cartan forms as in (\ref{vU=UOm}).
This is the case because the  variation of spinorial
harmonics are expressed through the same Cartan forms,
\begin{eqnarray}\label{vV=UdUG}
 V^{-1} \delta V = {1\over 4}
i_\delta\Omega^{(\underline{a})(\underline{b})} \,
\Gamma_{(\underline{a})(\underline{b})} \equiv {1\over 4}
(U^{-1}\delta U)^{(\underline{a})(\underline{b})} \,
\Gamma_{(\underline{a})(\underline{b})}  \; .   \qquad
\end{eqnarray}
This reflects the fact that locally the spinorial harmonics carry
the same $D(D-1)/2$ degrees of freedom as the vector ones, which is
tantamount to stating that the groups $Spin(1,D-1)$ and $SO(1,D-1)$
are locally isomorphic. In this sense {\it  moving frame action can
always be considered as spinor moving frame action}.

\subsection{Generalized action principle\label{secGAP}}

Generalized action principle for superbranes \cite{bsv} gives an
extended object counterpart of the rheonomic approach to
supergravity \cite{Regge,rheoB}. It can be obtained from the spinor
moving frame action by the following two steps.

{\it First} one replaces all the worldvolume {\it fields} dependent
on $W^{(p+1)}$ local coordinates $\xi^m$ by superfields, depending
on the local bosonic and fermionic coordinates, $\xi^m$ and
$\eta^q$, of the worldvolume superspace $W^{(p+1|n/2)}$,
\begin{eqnarray}\label{Vspf}
\hat{Z}^{{M}}(\xi) \mapsto \hat{Z}^{{M}}(\xi,\eta )\; , \quad V(\xi)
\mapsto V(\xi,\eta )\quad \Rightarrow \quad
u_{\underline{b}}{}^a(\xi) \mapsto u_{\underline{b}}{}^a(\xi,\eta
)\; . \qquad
\end{eqnarray}
{\it Secondly}, one replaces the integral over worldvolume
$W^{(p+1)}$ by integral over a surface $\widetilde{W}{}^{(p+1)}$ of
maximal bosonic dimension in the worldvolume superspace. Its
embedding into $W^{(p+1|n/2)}$ can be described by fermionic
coordinate functions $\hat{\eta}^q(\xi)$, which provide the
counterparts of the Volkov-Akulov Goldstone fermions (these would be
$\theta^{\check{\alpha}}(x)$ in our notation) \cite{VA72,VS73},
\begin{eqnarray}\label{MinW}
\widetilde{W}{}^{(p+1)}\in W^{(p+1|n/2)}\; : \qquad
\eta^q=\hat{\eta}^q(\xi)\; .
\end{eqnarray}
This is tantamount to saying that the generalized action is given by
Eq. (\ref{Sp=sp-mfr}) with an integral over the bosonic body of the
worldvolume superspace $W^{(p+1|n/2)}$, which is defined by
$\hat{\eta}^{{\check{q}}}=0$ and is denoted by $W^{(p+1)}$, and with
the Lagrangian form $\hat{E}^{\wedge (p+1)}- p \hat{B}_{p+1}$
constructed from superfields (\ref{Vspf}) pulled back to the surface
$\widetilde{W}{}^{(p+1)}$ in the worldvolume superspace, {\it i.e.}
from
\begin{eqnarray}\label{VspfM}
\hat{Z}^{{M}}=\hat{Z}^{{M}}(\xi,\hat{\eta}(\xi) )\; , \quad V=
V(\xi,\hat{\eta}(\xi) )\quad \Rightarrow \quad
u_{\underline{b}}{}^a=u_{\underline{b}}{}^a(\xi,\hat{\eta}(\xi) )\;
,
\\ \label{Ea:=GAP}
\hat{E}^{a} := {E}^{\underline{b}}(\hat{Z}(\xi,\hat{\eta}(\xi)))\;
u_{\underline{b}}{}^a(\xi,\hat{\eta}(\xi)) \; . \qquad
\end{eqnarray}

To resume, the {\it generalized action functional} is given by
\begin{eqnarray}\label{Sp=sp-GAP}
  S_p & =  \int\limits_{\widetilde{W}{}^{p+1}} (\hat{E}^{\wedge ({p+1})} -
  p  \hat{B}_{p+1}) :=
   \int\limits_{W^{p+1}} (\hat{E}^{\wedge ({p+1})}  -
  p  \hat{B}_{p+1})\vert_{\eta=\hat{\eta}(\xi)}
   \; ,   \qquad
 \end{eqnarray}
where hat ($\hat{\;}$) implies pull--back to the worldvolume
superspace, and also  (spinor) moving frame variables are
superfields, as in (\ref{Vspf}). Thus the original moving frame
action (\ref{Sp=sp-mfr}) is a particular case of the generalized
action for $\widetilde{W}{}^{p+1}=W^{p+1}$, {\it i.e.} for
$\hat{\eta}^q(x)=0$.

The set of equations of motion for this generalized action
functional includes a counterpart of (\ref{Ei=0}), but for the
superfields pulled back to $\widetilde{W}{}^{p+1}$,
\begin{equation}\label{Ei=0GAP}
\hat{E}^i:= {E}^{\underline{b}}(\hat{Z}(\xi,\hat{\eta}(\xi)))\,
u_{\underline{b}}^{\;\; i} (\xi , \hat{\eta}(\xi)) = 0 \; ,
\end{equation}
and also the dynamical equations of motion (\ref{Omaai=}),
(\ref{fEq-p}) but for the superfields pulled back to $\widetilde{W}{}^{p+1}$.

Now the structure of the Lagrangian form guaranties that the action
functional is independent on the choice of the surface $\widetilde{W}{}^{p+1}$.
The arbitrary changes of this surface, which are described by
arbitrary variations of the fermionic functions $\delta
\hat{\eta}^q(\xi)$ are the gauge symmetry of the generalized action
functional (\ref{Sp=sp-GAP}). More details on this symmetry can be
found in \cite{bsv} as well as in very recent
\cite{Berkovits:2008qw} which uses a 'bottom-up' version of the
generalized action principle proposed in \cite{Howe-GAP}. The
consequence of this symmetry `parametrized' by arbitrary
$\delta\eta^q(\xi)$'s is that equations of motion, including
(\ref{Ei=0GAP}) are valid {\it on an arbitrary surface $\widetilde{W}{}^{p+1}$} in
the worldvolume superspace $W^{(p+1|n/2)}$. As the set of such
surfaces 'covers' the whole superspace $W^{(p+1|n/2)}$, it is {\it
natural to assume} that the equations  are valid in the whole
superspace. This implies, in particular, lifting of  Eq.
(\ref{Ei=0GAP}) to the superembedding equation in its form of Eq.
(\ref{sEi=0}),
\begin{equation}\label{sEi=0GAP}
\hat{E}^i= E^{\underline{b}}(\hat{Z}(\xi,\eta))
u_{\underline{b}}^{\;\; i} (\xi,\eta) = 0 \; . \qquad
\end{equation}

This last stage, namely the {\it lifting} of the equations valid on
an arbitrary surface in superspace to equations on the superspace,
is the essence of the {\it rheonomic principle} of the group
manifold approach to supergravity\cite{Regge,rheoB}.

Notice that such a rheonomic lifting does not follow from the action
variation, but rather constitutes an additional stage in the
procedure of the {\it generalized action principle}, which should be
made separately after varying the generalized action {\it
functional}. In particular, the lifted equations written in terms of
complete superfields (not pulled back to $M^{p+1}$) should be
checked on consistency, and the consistency is not guaranteed. It
have to be checked case by case, see \cite{bpst} for an example when
the consistency does not hold.

The study of the selfconsistency condition for superembedding
equation (\ref{sEi=0GAP}) can actually be used to derive equations
of motion for D=10 type  superstrings, D=11
supermembranes\cite{bpstv} as well as for M5-brane\cite{hs2} and
D=10 super-D$p$-branes \cite{hs1} for $p\leq 5$\cite{hs+}. In the
next section  we will show explicitly how this happens in the case
of superstring in general curved type IIB supergravity superspace.

\section{Superembedding approach
to $D=10$ Green--Schwarz superstring in type IIB supergravity
background.} \label{sec1.3}

To discuss the superembedding approach to  superstring in a general type IIB supergravity background,  we need firstly to discuss the specific features of
the stringy spinor moving frame formalism.

\subsection{Spinor moving frame action for superstring}
\label{sec1.3.2}

The special properties of the stringy (spinor) moving frame
variables, {\it i.e.} of the ${SO(1,D-1)\over SO(1,1)\otimes
SO(D-2)}$ harmonics used to describe the $D$ dimensional
superstring, comes from the fact that the two dimensional $SO(1,1)$
pseudo rotations of the vectors $u_{\underline{a}}^a=
(u_{\underline{a}}^0\, ,\, u_{\underline{a}}^{\#} )$ (where the
symbol $\#$ is used for $(D-1)$-the direction) is reducible and can
be split onto the scaling of two light--like vectors $u_m^{++}:=
u_m^0 + u_m^{\#}$ and $u_m^{--}:= u_m^0 - u_m^{\#}$ (the self-dual
and anti-selfdual 2--vectors) by mutually inverse factors,
$u_m^{\pm\pm}\mapsto e^{\pm 2\alpha}u_m^{\pm\pm} $. The constraints
on the stringy moving frame variables (vector
 harmonics)
\begin{eqnarray}\label{harmU10in}
U_{\underline{a}}^{(\underline{b})}= (u_{\underline{a}}^{--},
u_{\underline{a}}^{++}, u_{\underline{a}}^{j})\; \in \; SO(1,9)
\qquad
\end{eqnarray}
read ({\it cf.} (\ref{UTIU=I}), (\ref{uIu=I}))
\begin{eqnarray}\label{harmU10}
 U^T\eta U = \eta & \qquad \Leftrightarrow \qquad \begin{cases}
u_{\underline{a}}^{--}u^{{\underline{a}}--}=0 \; , \quad
u_{\underline{a}}^{++}u^{{\underline{a}}++}=0 \; , \cr
u_{\underline{a}}^{--}u^{{\underline{a}}++}=2 \; , \quad
u_{\underline{a}}^{\pm\pm}u^{{\underline{a}}\, i}=0 \; , \quad \cr
u_{\underline{a}}^{i}u^{{\underline{a}}\, j}=- \delta^{ij}
\end{cases}\;  \qquad \end{eqnarray}
and also imply
\begin{eqnarray}
\label{UIUT=str}
 & U\eta U^T = \eta  \qquad \Leftrightarrow \qquad
\delta_{\underline{a}}{}^{\underline{b}}= {1\over
2}u_{\underline{a}}^{++}u^{\underline{b}--} + {1\over
2}u_{\underline{a}}^{--}u^{{\underline{b}}++} -
u_{\underline{a}}^{i}u^{{\underline{b}} i} \qquad
\end{eqnarray}
Then, induced worldvolume supervielbein (\ref{Ea:=EuaP}) are
\begin{eqnarray}\label{E++E--=}
\hat{E}^{++}:= \hat{E}^{\underline{a}}u_{\underline{a}}^{++}\; ,
\qquad \hat{E}^{--}:=
\hat{E}^{\underline{a}}u_{\underline{a}}^{--}\; \qquad
\end{eqnarray}
and the spinor moving frame action for superstring reads (see \cite{IB2001})
\begin{equation}\label{SIIB-2ord}
S_{IIB}=   {1 \over 2} \int_{W^{2}}\;  \hat{E}^{++} \wedge
\hat{E}^{--} - \int_{W^{2}} \hat{B}_2 \; ,
\end{equation}
or, using the auxiliary worldvolume vielbein forms ${e}^{\pm\pm}$  (see \cite{BZ-str}), as
\begin{eqnarray}\label{SIIB-1ord}
S_{IIB} & = {1 \over 2} \int_{W^{2}} \left(  {e}^{++} \wedge
\hat{E}^{--} -   {e}^{--} \wedge \hat{E}^{++}
 - {e}^{++} \wedge
{e}^{--} \right) - \int_{W^{2}} \hat{B}_2 \; .\qquad
\end{eqnarray}
Indeed, $\delta {e}^{\pm\pm}$ equations of motion express them
through  $\hat{E}^{\pm\pm}$ of (\ref{E++E--=}),
\begin{eqnarray}\label{e++=E++} & e^{\pm\pm}=  \hat{E}^{\pm\pm}:=
\hat{E}^{\underline{a}}u_{\underline{a}}^{\pm\pm}\; . \qquad
\end{eqnarray}
 Substituting the algebraic equations (\ref{e++=E++}) back to the first order action (\ref{SIIB-1ord}) one
arrives at the second order  action  (\ref{SIIB-2ord}).

As we discussed in Sec.  \ref{secGAP},
the above spinor moving frame action can be used to construct the
generalized action\cite{bsv}. This is given by formally the same
functional (\ref{SIIB-2ord}) (or (\ref{SIIB-1ord})) with the fields
on ${W}{}^2$ replaced by the superfields and integration performed about
an arbitrary surface $\widetilde{W}{}^2$ in the worldsheet superspace
$W^{(2|8+8)}$. The generalized action principle for superstring
produces in particular, the superembedding equation (\ref{sEi=0})
\cite{bsv}.

\subsection{Stringy ${Spin(1,9)\over SO(1,1)\otimes SO(8) }$ spinorial harmonics}

The D$=$10 stringy spinor harmonics are collected in $Spin(1,9)$
matrix
\begin{eqnarray}\label{VSTRin}
V_\alpha^{(\beta)}= (v_\alpha{}_q^{+}\; ,
v_\alpha{}_{\dot{q}}^{-})\; \in \; Spin(1,9) \; . \qquad
\end{eqnarray}
The specific of string lays in that the spinor representation of
$SO(1,1)$ is one dimensional and is described by sign indices $^+$
and $^-$ of $v_\alpha{}_q^{-}$ and $v_\alpha{}_{\dot{q}}^{+}$. For
our  D$=$10 case  $q$ and $\dot{q}$  are the $s$- and $c$-spinorial
indices of $SO(8)$.

In the dynamical system with $SO(1,1)\otimes SO(8)$ symmetry, like
our superstring described by the action (\ref{SIIB-1ord}) or
(\ref{SIIB-2ord}), the harmonics are homogeneous coordinates of the
coset ${Spin(1,9)\over SO(1,1)\otimes SO(8) }$,
\begin{eqnarray}\label{VSTRinG-H}
 \{ V_\alpha^{(\beta)}\} = \{ (v_{\alpha q}^{\;\,+}\; ,
v_{\alpha\dot q}^{\;\,-})\} \; = \; {Spin(1,9)\over SO(1,1)\otimes
SO(8) }\; . \qquad
\end{eqnarray}
The requirement for the matrix $V$ to belong to $Spin(1,9)$ group
(\ref{VSTRin}) is imposed as the (reducible) constraint
\begin{eqnarray}\label{vssSTR}
\sigma^{\underline{b}}U^{(\underline{a})}_{\underline{b}} = V
\sigma^{(\underline{a})} V^T\; , \qquad (a) \qquad V^T
\tilde{\sigma}_{\underline{b}} V=U^{\,
(\underline{a})}_{\underline{b}} \tilde{\sigma}_{(\underline{a})} \;
, \qquad (b) \;  \qquad
\end{eqnarray}
which express the Lorentz invariance of the D$=$10 sigma--matrices
${\sigma}^{\underline{b}}_{{\alpha}{\beta}}$,
$\tilde{\sigma}_{\underline{b}}^{{\alpha}{\beta}}$. These are
symmetric, obey
${\sigma}^{\underline{a}}\tilde{\sigma}_{\underline{b}}+
{\sigma}^{\underline{a}}\tilde{\sigma}_{\underline{b}}=\delta_\alpha{}^\beta$
and have  $Spin(1,1) \otimes SO(8)$ invariant representation with
which the constraints (\ref{vssSTR}a) can be split into the
following set of relations
 \begin{eqnarray}\label{u++=v+v+2}
 u^{++}_{\underline{a}}
\sigma^{\underline{a}}_{{\alpha}{\beta}} = 2v_{{\alpha}q}^{~+}
v^{~+}_{{\beta}q}\; , & \qquad &
 u^{++}_{\underline{a}}
\tilde{\sigma}^{\underline{a}{\alpha}{\beta}} =
2v^{+{\alpha}}_{\dot{q}} v_{\dot{q}}^{+{\beta}},  \qquad
\\ \label{u--=v-v-2}
 u^{--}_{\underline{a}}
\sigma^{\underline{a}}_{{\alpha}{\beta}} = 2v_{{\alpha}\dot{q}}^{~-}
v^{~-}_{{\beta}\dot{q}}\; , & \qquad &
 u^{--}_{\underline{a}}
\tilde{\sigma}^{\underline{a}{\gamma}{\beta}} = 2v^{-{\gamma}}_{q}
v_{q}^{-{\beta}}\; , \qquad
\\ \label{ui=v+v-2}
 u^{i}_{\underline{a}}
\sigma^{\underline{a}}_{{\alpha}{\beta}} = 2v_{({\alpha}{q}}^{~+}
\gamma^i_{q\dot{q}} v^{~-}_{{\beta})\dot{q}}\; , & \qquad &
 u^{i}_{\underline{a}}
\tilde{\sigma}^{\underline{a}{\gamma}{\beta}} = -
v^{-{\gamma}}_{{q}} \gamma^i_{q\dot{q}} v_{\dot{q}}^{+{\beta}} -
v^{-{\beta}}_{{q}} \gamma^i_{q\dot{q}} v_{\dot{q}}^{+{\gamma}}\; .
\qquad
\end{eqnarray}
These imply, in particular, that the spinor harmonics $
v^{+{q}}_{{\alpha}}, v^{-\dot{q}}_{{\alpha}}$ can be treated as
square roots from the light--like vectors $u^{++}_{\underline{a}},
u^{--}_{\underline{a}}$.

The second relations in (\ref{u++=v+v+2})-(\ref{ui=v+v-2}) are
written for  inverse harmonics
\begin{eqnarray}\label{V-1STR}
 V_{(\alpha)}{}^{\beta} = (v^{- {\beta}}_q\; , v^{+ {\beta}}_{\dot{q}} ) \; \in \;
Spin(1,9)\; .
\end{eqnarray}
In the case of $D=10$ Majorana--Weyl spinor representation,
with $\alpha=1,\ldots , 16$, these cannot be constructed from the
`original' spinorial harmonics (\ref{VSTRinG-H}) and are defined by
the constraints
\begin{eqnarray}\label{V-1=}
 V_{(\alpha)}^{\;\;\;
\gamma} V_{\gamma}^{(\beta)} =\delta_{(\alpha)}^{\;\; (\beta)}:=
\left(\begin{matrix} \delta_{{q}}^{{p}} & 0\cr 0 &
\delta_{\dot{q}}^{\dot{p}} \end{matrix}\right)\;
\end{eqnarray}
(like {\it e.g.} inverse metric  in general relativity). Eq.
(\ref{V-1=}) implies
\begin{eqnarray}\label{V-1V=ISTR}
v^{-{\alpha}}_{p} v^{~+}_{{\alpha}q}= \delta_{pq}, \qquad
v^{-{\alpha}}_{p} v^{~-}_{{\alpha}\dot{q}} =0\, , \qquad \nonumber
\\
v^{+{\alpha}}_{\dot{p}} v^{~+}_{{\alpha}q}=0, \qquad
v^{+{\alpha}}_{\dot{p}} v^{~-}_{{\alpha}\dot{q}}
=\delta_{\dot{p}\dot{q}}\; . \qquad
\end{eqnarray}
These relations can be used to factorize the projector and to get
the irreducible form of the superstring $\kappa$--symmetry
\cite{BZ-str}. They are also necessary to develop the superembedding
approach to superstrings \cite{bpstv,IB2001}.

Finally, the split form of Eq. (\ref{vssSTR}b) reads
\begin{eqnarray}\label{v+sv+=u++}
v_{q}^{+}\tilde{\sigma}^{\underline{a}}v_{p}^{+}= \delta_{qp}
u^{++}_{\underline{a} }\; , & \qquad &
v_{\dot{q}}^{+}{\sigma}^{\underline{a}}v_{\dot{p}}^{+}=
\delta_{\dot{q}\dot{p}} u^{++}_{\underline{a} }\; , \qquad \\
\label{v-sv-=u--}
v_{\dot{q}}^{-}\tilde{\sigma}^{\underline{a}}v_{\dot{p}}^{-}=
\delta_{qp} u^{--}_{\underline{a}} \; , & \qquad &
v_{{q}}^{-}{\sigma}^{\underline{a}}v_{{p}}^{-}=
\delta_{\dot{q}\dot{p}} u^{--}_{\underline{a}} \; , \qquad
\\ \label{v-sv+=ui}
v_{{q}}^{+}\tilde{\sigma}^{\underline{a}}v_{\dot{q}}^{-}=
\gamma^i_{q\dot{q}}  u^{i}_{\underline{a}} \; , & \qquad &
v_{{q}}^{-}{\sigma}^{\underline{a}}v_{\dot{q}}^{+}= -
\gamma^i_{{q}\dot{q}} u^{i}_{\underline{a}} \; . \qquad
\end{eqnarray}

\subsection{Superembedding approach to D=10 superstring in
type IIB supergravity background }

The starting point is the superembedding equation in its form of Eq.
(\ref{sEi=0}),
\begin{equation}\label{Ei=0STR}
\hat{E}^i:= \hat{E}^{\underline{b}}u_{\underline{b}}^{\;\; i} (\xi
,\eta)= 0 \; .
\end{equation}
This has to be completed by the set of {\it conventional
constraints} which includes (\ref{e++=E++})  and the relations
defining fermionic supervielbein forms
\begin{eqnarray}\label{epm=Epm}  e^{\pm\pm}=  \hat{E}^{\pm\pm}:=
 \hat{E}^{\underline{a}}u_{\underline{a}}^{\pm\pm}\; ,
\qquad \\ \label{e+q=e-q=} \qquad e^{+q}= \hat{E}^{\alpha 1} v^{\;\;
+}_{\alpha q}\; , \qquad e^{-\dot{q}}= \hat{E}^{\alpha 2} v^{\;\;
-}_{\alpha \dot{q}}\; .
\end{eqnarray}

The superembedding equation (\ref{Ei=0STR}) and the above set of
conventional constraints can be collected in the following
expressions for the pull--back of the supervielbein of target type
IIB superspace,
\begin{eqnarray}\label{hEa=STR}
&& \hat{E}^{\underline{a}}:= {1\over 2}e^{++} u^{--\underline{a}} +
{1\over 2}e^{--} u^{++\underline{a}} \; ,
\\ \label{hEf12=STR}
 && \hat{E}^{\alpha 1} = e^{+q} v^{-\alpha}_{q} + e^{\pm\pm}
\chi^{-\dot{q}}_{_{\pm\pm}}  v^{+\alpha}_{\dot{q}}\; , \qquad
 \hat{E}^{\alpha 2} = e^{-\dot{q}} v^{+\alpha}_{\dot{q}}  + e^{\pm\pm}
\chi^{+{q}}_{_{\pm\pm}}  v^{-\alpha}_{q}\; .  \qquad
\end{eqnarray}
Actually, Eqs. (\ref{hEf12=STR}) contain a bit more than just Eq.
(\ref{e+q=e-q=}): it also states that  $\hat{E}_{+p}^{\; \alpha 1}
v^{\;\; -}_{\alpha \dot{q}}= 0$ and $\hat{E}^{\; \alpha
2}_{-\dot{p}} v^{\;\; +}_{\alpha q}= 0$, and this excludes from
consideration the case of D1-branes (see \cite{bst,IB2001}).

\subsubsection{Other conventional constraints}

To complete the set of conventional constraints, let us notice that
we use the  $SO(1,1) \otimes SO(8)$ connection induced by embedding;
this implies that the complete $SO(1,9) \otimes SO(1,1) \otimes
SO(8)$  covariant derivatives of the vector harmonics read
\begin{eqnarray}\label{Du+-i=STR}
& \begin{cases} Du^{++}_{\underline{a}} =
u^{i}_{\underline{a}}\Omega^{++ i} \; , \cr Du^{--}_{\underline{a}}
= u^{i}_{\underline{a}}\Omega^{-- i} \; , \end{cases}\qquad
Du^{i}_{\underline{a}} = {1\over 2}u^{--}_{\underline{a}}\Omega^{++
i} + {1\over 2}u^{++}_{\underline{a}}\Omega^{-- i} \; .  \qquad
\end{eqnarray}
For the spinorial harmonics (\ref{VSTRinG-H}), (\ref{V-1STR}) this
connection gives
\begin{eqnarray}\label{Dv+q=STR}
\qquad & Dv_{\alpha {q}}^{\;\;  +}= {1\over 2} \Omega^{++i}
\gamma_{q\dot{p}}^{i} v_{\alpha \dot{p}}^{\;\; +} \; , \qquad
Dv_{\dot{q}}^{+\alpha}= -{1\over 2}\Omega^{++i}\;
v_{{p}}^{-\alpha}\gamma^i_{p\dot{q}}
\; , \qquad \\
\label{Dv-dq=STR} & Dv_{\alpha \dot{q}}^{\;\;  -}= {1\over 2}
\Omega^{--i} \; v_{\alpha {p}}^{+}\gamma_{p\dot{q}}^{i} \; ,  \qquad
Dv_{{q}}^{-\alpha}= -{1\over 2}\Omega^{--i}\; \gamma^i_{q\dot{p}}
v_{\dot{p}}^{+\alpha}\; .  \qquad
\end{eqnarray}

The integrability conditions for Eqs. (\ref{Du+-i=STR}) give the
curved superspace generalization of the Peterson-Codazzi,  Gauss and
Ricci equations of the classical XIX-th century surface theory.
These read
\begin{eqnarray}\label{DOm+-i=STR}
 D\Omega^{\pm\pm i} &=& \hat{R}^{\pm\pm i}:=
\hat{R}^{\underline{a}\underline{b}}u^{\pm\pm }_{\underline{a}}u^{i
}_{\underline{b}} \; , \qquad  \\ \label{DOm0=STR}  d\Omega^{(0)} \;
&=& {1\over 4} \hat{R}^{\underline{a}\underline{b}}u^{++
}_{\underline{a}}u^{-- }_{\underline{b}} + {1\over 4}  \Omega^{--
i}\wedge \Omega^{++ i}\; , \qquad  \\ \label{DOmij=STR}
\mathbb{R}^{ij} \; &=& \hat{R}^{\underline{a}\underline{b}}u^{i
}_{\underline{a}}u^{j}_{\underline{b}}\, - \,\Omega^{-- [i}\wedge
\Omega^{++ j]} \; ,  \qquad
\end{eqnarray}
where $\Omega^{\pm\pm i}$ are the generalized Cartan forms
(see Sec. \ref{CartanF}),
\begin{eqnarray}\label{Om++i=}
\Omega^{\pm\pm\; i}:= u^{\pm\pm\;
\underline{a}}\left(du^{i}_{\underline{a}} +
\omega_{\underline{a}}{}^{\underline{b}} u^{i}_{\underline{b}}
\right)\; ,
\end{eqnarray}
$\Omega^{(0)}= {1\over
4}u^{\underline{a}--}((d+\hat{w})u^{++})_{\underline{a}}$ is the
$SO(1,1)$ connection (the induced 2d spin connection) and
 $\mathbb{R}^{ij} =d\Omega^{ij}-\Omega^{ik}\wedge
\Omega^{kj}$ is the curvature of normal bundle with $\Omega^{ij}=
u^{\underline{a}i}((d+\hat{w})u^{j})_{\underline{a}}$; finally,
$\hat{w}_{\underline{a}}{}^{\underline{b}}$ is the pull back of the
$D=10$ spin connection superform
$\hat{w}_{\underline{a}}{}^{\underline{b}}= dZ^M{w}_{M,
\underline{a}}{}^{\underline{b}}$ and
$R^{\underline{a}\underline{b}}= (dw-w \wedge
w)^{\underline{a}\underline{b}}$.

\subsubsection{Torsion constraints}
Below we will also need the type IIB torsion constraints,
\begin{eqnarray}\label{Ta=consSTR}
{T}^{\underline{a}} & =  -i  {\cal E}^{\underline{\alpha}} \wedge
{\cal E}^{\underline{\beta}}
\Gamma^{\underline{a}}_{\underline{\alpha}\underline{\beta}} \; := -
i {E}^{\alpha 1} \wedge {E}^{\beta 1}
\sigma^{\underline{a}}_{\alpha\beta} - i {E}^{\alpha 2} \wedge
{E}^{\beta 2} \sigma^{\underline{a}}_{\alpha\beta}\; , \qquad
\\ \label{Tal1=str}
 T^{\alpha 1} &= -E^{\alpha 1}\wedge E^{\beta 1} \nabla_{\beta 1}e^{-{\Phi}} + {1\over
2} E^{1}\sigma^{\underline{a}}\wedge E^{1}\,
\tilde{\sigma}_{\underline{a}}^{\alpha\beta} \nabla_{\beta
1}e^{-{\Phi}} + \qquad \nonumber \\ & \qquad +
E^{\underline{a}}\wedge {\cal E}^{{\underline{\beta}}}
T_{{\underline{\beta}}\underline{a}}{}^{\alpha 1 }
 +  {1\over 2}E^{\underline{b}} \wedge E^{\underline{a}}
 T_{\underline{a}\underline{b}}{}^{\alpha 1 }  \; ,  \qquad
\\
 \label{Tal2=str}  T^{\alpha 2} &= -E^{\alpha 2}\wedge E^{\beta 2}
\nabla_{\beta 2}e^{-{\Phi}} + {1\over 2}
E^{2}\sigma^{\underline{a}}\wedge E^{2}\,
\tilde{\sigma}_{\underline{a}}^{\alpha\beta} \nabla_{\beta
2}e^{-{\Phi}} + \qquad \nonumber \\ & \qquad +
E^{\underline{a}}\wedge {\cal E}^{{\underline{\beta}}}
T_{{\underline{\beta}}\underline{a}}{}^{\alpha 2 }
 +  {1\over 2}E^{\underline{b}} \wedge E^{\underline{a}}
 T_{\underline{a}\underline{b}}{}^{\alpha 2 }  \; ,  \qquad
\\ \label{cEf=IIBSTR}   {\cal E}^{\underline{\alpha}} &= ( E^{\alpha
\, 1 } \; ,\; E^{\alpha \, 2} ) \; , \quad ^{\quad\alpha =1,\ldots , 16\;
,}_{\underline{\alpha}=(\alpha i)=1,\ldots , 32\; ,} \qquad
\Gamma^{\underline{a}}_{\underline{\alpha}\underline{\beta}}= diag
\left(\sigma^{\underline{a}}_{\alpha\beta}\; , \;
\sigma^{\underline{a}}_{\alpha\beta}\right) \, .  \qquad
\end{eqnarray}
The  fermionic torsions
$T_{\underline{\beta}\underline{a}}{}^{\underline{\alpha}
}=(T_{{\underline{\beta}}\underline{a}}{}^{\alpha 1},
T_{{\underline{\beta}}\underline{a}}{}^{\alpha 2 })$ can be read off from
\begin{eqnarray}\label{EThfbhf=}
{\cal E}^{{\underline{\alpha}}}
T_{{\underline{\alpha}}\underline{b}}{}^{{\underline{\beta}} } & = -
{1\over 8} \left( H_{\underline{b}\underline{c}\underline{d}}\,
\sigma^{\underline{c}\underline{d}}\tau_3 + \sigma_{\underline{b}}
\tilde{R}\!\!\!/{}^{(1)}i\tau_2 - \sigma_{\underline{b}}
\tilde{R}\!\!\!/{}^{(3)}\tau_1 + -\!\!\!\! ^1_2 \;
\sigma_{\underline{b}} \tilde{R}\!\!\!/{}^{(5)}i\tau_2
\right){}_{\hat{\underline{\alpha}}}{}^{\hat{\underline{\gamma}}}
\qquad \\ \nonumber  &= - {1\over 8} {\cal E}^{\underline{\alpha}}
H_{\underline{b}\underline{c}\underline{d}}\, (i\tau_3 \otimes
\sigma^{\underline{c}\underline{d}})_{{\underline{\alpha}}}{}^{{\underline{\beta}}}\;
 +  {1\over 16} {\cal E}^{\underline{\alpha}}\, \sum\limits_{n=0}^{4}
(\sigma_{\underline{b}} \tilde{R}\!\!\!/{}^{(2n+1)}\otimes
\tau_1(\tau_3)^n )_{{\underline{\alpha}}}{}^{{\underline{\beta}}} \;
. \;
\end{eqnarray}
Here $\tau_3\sigma^{\underline{c}\underline{d}}= \tau_3 \otimes
\sigma^{\underline{c}\underline{d}}$, etc. and
$\tilde{R}\!\!\!/{}^{(2n+1)}= {1\over (2n+1)!}\,
R_{\underline{c}_1\ldots \underline{c}_{2n+1}}
\tilde{\sigma}^{\underline{c}_1\ldots \underline{c}_{2n+1}\;
\alpha\beta}$ where  $R_{\underline{a}_1 \ldots
\underline{a}_{9-2n}}$ are the type IIB RR field strength. Notice
that $R_{\underline{a}_1 \ldots \underline{a}_{9-2n}} = {(-)^n \over
(2n+1)!}\; \varepsilon_{\; \underline{a}_1 \ldots
\underline{a}_{9-2n}\underline{b}_1 \ldots \underline{b}_{2n+1}}
R^{\underline{b}_1 \ldots \underline{b}_{2n+1}}$, which describes, in
particular, the self-duality of the 5-form field strength.

\subsubsection{Superstring equations of motion from superembedding}

The selfconsistency conditions for the superembedding equations is
included in the integrability condition of Eqs. (\ref{hEa=STR}),
\begin{eqnarray}\label{DhEua=STR}
&& \hat{T}^{\underline{a}}  =   - i \hat{E}^{\alpha 1} \wedge
\hat{E}^{\beta 1} \sigma^{\underline{a}}_{\alpha\beta} - i
\hat{E}^{\alpha 2} \wedge \hat{E}^{\beta 2}
\sigma^{\underline{a}}_{\alpha\beta} = \qquad \nonumber \\ && \; =
{1\over 2}De^{++}u^{--}_{\underline{a}} + {1\over
2}De^{--}u^{++}_{\underline{a}} + {1\over 2} u^{\;
i}_{\underline{a}}(e^{++}\wedge \Omega^{-- i} + e^{--}\wedge
\Omega^{++ i} ) \; . \qquad
\end{eqnarray}
Using (\ref{hEf12=STR}) one finds, after some algebra, that the
contraction of Eq. (\ref{DhEua=STR}) with the light--like vectors
$u^{\pm\pm}_{\underline{a}}$ determine the worldvolume bosonic
torsion,
\begin{eqnarray}\label{De++=STR}
De^{\!^{++}}=-2i e^{+q}\wedge e^{+q} - 4i e^{\!^{++}}\wedge
e^{\!^{--}} \chi_{_{++}}^{+{q}}\chi_{_{--}}^{+{q}}\; , \qquad \\
\label{De--=STR} De^{\!^{--}}=-2i e^{-\dot{q}}\wedge e^{-\dot{q}} -
4i e^{\!^{++}}\wedge e^{\!^{--}}
\chi_{_{++}}^{-\dot{q}}\chi_{_{--}}^{-\dot{q}}\; ,
  \qquad
\end{eqnarray}
while the contraction with $u^{\;\; i}_{\underline{a}}$ gives the
restriction for the covariant Cartan forms (\ref{Om++i=}),
\begin{eqnarray}\label{e+Om-+=STR}
e^{\!^{++}}\wedge \Omega^{\!^{--i}} + e^{\!^{--}}\wedge
\Omega^{\!^{++i}}=-4i\gamma^i_{q \dot{q}} e^{\!^{\pm\pm}}\wedge
e^{+q} \chi_{_{\pm\pm}}^{-\dot{q}} -4i\gamma^i_{q \dot{q}}
e^{\!^{\pm\pm}}\wedge e^{-\dot{q}} \chi_{_{\pm\pm}}^{+q}\; . \qquad
\end{eqnarray}

To proceed further one needs to study the consistency
(intergability) of the fermionic conventional constraints
(\ref{hEf12=STR}) which read
\begin{eqnarray}\label{DEf1=STR}
 \hat{T}^{\alpha 1} &=&  De^{+q} v^{-\alpha}_{q} + e^{+q} \wedge Dv^{-\alpha}_{q} + De^{\pm\pm}
\chi^{-\dot{q}}_{_{\pm\pm}}  v^{+\alpha}_{\dot{q}} +
e^{\pm\pm}\wedge D\chi^{-\dot{q}}_{_{\pm\pm}}  v^{+\alpha}_{\dot{q}}
+ \qquad \nonumber \\ && \qquad + e^{\pm\pm}\wedge
\chi^{-\dot{q}}_{_{\pm\pm}}  Dv^{+\alpha}_{\dot{q}} \; , \qquad
\end{eqnarray}
\begin{eqnarray}
\label{DEf2=STR}
 \hat{T}^{\alpha 2} &=&  De^{-\dot{q}}v^{+\alpha}_{\dot{q}} + e^{-\dot{q}} \wedge Dv^{+\alpha}_{\dot{q}}+
  De^{\pm\pm} \chi^{+{q}}_{_{\pm\pm}}  v^{-\alpha}_{q}
 + e^{\pm\pm}\wedge
D\chi^{+{q}}_{_{\pm\pm}}  v^{-\alpha}_{q} + \qquad \nonumber \\ && +
e^{\pm\pm}\wedge \chi^{+{q}}_{_{\pm\pm}} Dv^{-\alpha}_{q}\; . \qquad
\end{eqnarray}
The right hand side of these equations can be specified by using
Eqs. (\ref{Dv+q=STR}), (\ref{Dv-dq=STR}). To specify the {\it
l.h.s.}'s we need the explicit form of the fermionic torsion
constraints for the type IIB superspace, Eqs. (\ref{Tal1=str}),
(\ref{Tal2=str}).

Contracting (\ref{DEf1=STR})  with $v_{\alpha q}^{\;\; +}$ and
(\ref{DEf2=STR}) with $v_{\alpha \dot{q}}^{\;\; -}$ one finds the
fermionic torsion of the worldvolume superspace. These read
\begin{eqnarray}\label{De+q=STR}
De^{+q}= - e^{+p} \wedge e^{+p^\prime} (\delta_{(p}^{\;\; q}
v_{p^\prime)}^{-\,\beta} - \delta_{pp^\prime}v_{q}^{-\,\beta})
\widehat{D_{\beta 1}e^{-\Phi}} + \propto e^{\pm\pm} \; , \\
\label{De-dq=STR} De^{-\dot{q}}= - e^{-\dot{p}} \wedge
e^{-\dot{p}^\prime} (\delta_{(\dot{p}}^{\;\; \dot{q}}
v_{\dot{p}^\prime )}^{+\,\beta} -
\delta_{\dot{p}\dot{p}^\prime}v_{\dot{q}}^{+\,\beta})
\widehat{D_{\beta 2}e^{-\Phi}} + \propto e^{\pm\pm} \; ,
\end{eqnarray}
where $\propto e^{\pm\pm}$ denotes the contributions from forms
containing bosonic worldvolume supervielbein, which we will not need
in this section.

Contracting (\ref{DEf1=STR}) with $v_{\alpha \dot{q}}^{\;\; -}$ one
arrives at
\begin{eqnarray}\label{Tf1v-dq=STR}
0 &=& -{1\over 2}e^{+p} \wedge
e^{+q}\Omega^{\!{--i}}_{+(q}\gamma^i_{p)\dot{q}} - 2i e^{+p} \wedge
e^{+p} \chi_{\!_{++}}^{\;\;-\dot{q}} -{1\over 2}e^{+q} \wedge
e^{-\dot{p}}\Omega^{\!{--i}}_{-\dot{p}}\gamma^i_{q\dot{q}} - \qquad
\nonumber \\ && \qquad - 2i e^{-\dot{p}} \wedge e^{-\dot{p}}
\chi_{\!_{--}}^{\;\;-\dot{q}} + \propto e^{\pm\pm} \; . \quad
\end{eqnarray}
The similar equation with $e^{+q}\leftrightarrow e^{-\dot{q}}$, $\pm
\leftrightarrow \mp$ appears when contracting (\ref{DEf2=STR}) with
$v_{\alpha {q}}^{\;\; +}$. An immediate consequence of these
equations are
\begin{eqnarray}\label{chi1/2=0}
\chi_{\!_{--}}^{\;\;-\dot{q}} := \hat{E}_{\!_{--}}^{\alpha 1}
v_{\alpha \dot{q}}^{\;\; -} = 0 \; ,  \qquad && \chi_{\!_{++}}^{\;\;
+{q}} := \hat{E}_{\!_{++}}^{\alpha 2} v_{\alpha {q}}^{\;\; +} = 0 \;
, \qquad  \\ \label{Om1/2=0} \Omega^{\!{--i}}_{-\dot{p}} =0  \; ,
\qquad && \Omega^{\!{++i}}_{+{p}} =0 \; . \quad
\end{eqnarray}
Eqs. (\ref{chi1/2=0}) are the equations of motion for the fermionic
degrees of freedom of superstring. A simple way to be convinced in
this is to observe that the linearized version of  (\ref{chi1/2=0})
can be written in the form similar to the light-cone gauge fermionic
equations of the Green-Schwarz superstring,  which define its 16
fermionic degrees of freedom as two chiral, namely  one right-moving
and one left-moving,  $8$ component fermions,
\begin{eqnarray}\label{d--Th-=0}
\partial_{--}\hat{\Theta}^{- 1}_{\dot{q}} = 0 \; , \qquad
\hat{\Theta}^{- 1}_{\dot{q}}(\xi, \eta):= \hat{\theta}^{\alpha 1}
v_{\alpha \dot{q} }^{\;\; -} \; , \qquad \\
\label{d++Th+=0}
\partial_{++}\hat{\Theta}^{+ 2}_{q} =0  \; , \qquad
\hat{\Theta}^{+ 2}_{{q}}(\xi, \eta):= \hat{\theta}^{\alpha 2}
v_{\alpha {q} }^{\;\; +} \; . \qquad
\end{eqnarray}
The above $\hat{\Theta}^{- 1}_{\dot{q}}$, $\hat{\Theta}^{+ 2}_{{q}}$
correspond to the light--cone gauge fermionic fields, but defined
with the use of {\it moving}  frame determined by the harmonics. The
other half of the fermionic fields, $\hat{\Theta}^{+ 1}_{{q}}:=
\hat{\theta}^{\alpha 1} v_{\alpha {q} }^{\;\; +}$ and
$\hat{\Theta}^{- 2}_{\dot{q}}:= \hat{\theta}^{\alpha 2} v_{\alpha
\dot{q} }^{\;\; -}$  can be identified with the the fermionic
coordinates of $W^{(2|8+8)}$,
\begin{eqnarray}\label{SEmGauge} \hat{\Theta}^{+
1}_{{q}}(\xi, \eta)=\eta^{+q}\;, \qquad  \hat{\Theta}^{-
2}_{\dot{q}}(\xi, \eta ) =\eta^{-\dot{q}}\;. \end{eqnarray}
These superfield relations imply  $\hat{\Theta}^{+
1}_{{q}}(\xi , 0) =0$, $\hat{\Theta}^{- 2}_{\dot{q}}(\xi , 0) =0$,
and these equations give a covariant version of the conditions which
might be fixed using the $\kappa$--symmetry of the standard
Green--Schwarz action.

\subsubsection{Bosonic equations of motion and
on-shell superembedding of the worldvolume
superspace}\label{sec1.3.4}

The fermionic equations of motion (\ref{chi1/2=0}) simplify the
expressions (\ref{hEf12=STR}) for the pull-back of fermionic
supervielbein forms, making them chiral,
\begin{eqnarray} \label{hEf1f2=STR}
 && \hat{E}^{\alpha 1} = e^{+q} v^{-\alpha}_{q} + e^{++}
\chi^{-\dot{q}}_{_{++}}  v^{+\alpha}_{\dot{q}}\; , \qquad
 \hat{E}^{\alpha 2} = e^{-\dot{q}} v^{+\alpha}_{\dot{q}}  + e^{--}
\chi^{+{q}}_{_{--}}  v^{-\alpha}_{q}\; ,  \qquad
\end{eqnarray}
which means, in particular, left-- and right--moving, but also
containing the corresponding half of the fermionic coordinates. The
bosonic worldsheet torsion (\ref{De++=STR}), (\ref{De--=STR})  and
Eq. (\ref{e+Om-+=STR})  also simplify,
\begin{eqnarray}\label{De+-=STR}
De^{\!^{++}}=-2i e^{+q}\wedge e^{+q}\; , \qquad De^{\!^{--}}=-2i
e^{-\dot{q}}\wedge e^{-\dot{q}} \; ,
  \qquad
\end{eqnarray}
and Eq. (\ref{e+Om-+=STR}) determines the generalized Cartan forms
to be
\begin{eqnarray}\label{Om--iSTR=}
\Omega^{\!^{--i}} = - 4i e^{+q} \gamma^i_{q\dot{q}}
\chi_{\!_{++}}^{-\dot{q}} +
e^{\!^{++}} \Omega^{^{--i}}_{\!_{++}}+e^{\!^{--}} K^i  \; , \qquad \\
\label{Om++iSTR=} \Omega^{\!^{++i}}= - 4i e^{-\dot{q}}
\chi_{\!_{--}}^{+{q}} \gamma^i_{q\dot{q}} +e^{\!^{++}} K^i   +
e^{\!^{--}} \Omega^{^{++i}}_{\!_{--}}  \; . \qquad
\end{eqnarray}
Using the  conventional constraints (\ref{e++=E++}), one can write
the mean curvature $\Omega^{^{--i}}_{\!_{--}}=
\Omega^{^{++i}}_{\!_{++}} := K^i$ in the form
\begin{eqnarray} \label{Ki:=DE} &  K^i := -
2D_{--}E_{++}^{\underline{b}}\, u^i_{\underline{b}}= -
2D_{++}E_{--}^{\underline{b}}\, u^i_{\underline{b}}\; . \qquad
\end{eqnarray}
Its linearized version reads $K^i= \Box X^i$, where $X^i=x^\mu
u_\mu^i$, so that one can expect the bosonic equations of motion to
appear in the form of conditions for $K^i$.

This is indeed the case. The bosonic equations of motion appears as
 $\propto e^{--}\wedge e^{+q}$ component of Eq. (\ref{Tf1v-dq=STR}).
First one obtains  $K^i\gamma^i_{q\dot{q}}= - {1\over 8}
u^{\underline{a}++}\hat{H}_{\underline{a}\underline{b}\underline{c}}
v_q^{-\alpha} \sigma^{\underline{b}\underline{c}}{}_\alpha{}^\beta
v_{\beta \dot{q}}^{\;\; -}$. Then, using Eqs. (\ref{v-sv-=u--}),
(\ref{v-sv+=ui}), one finds that $v_q^{-\alpha}
\sigma^{\underline{b}\underline{c}}{}_\alpha{}^\beta v_{\beta
\dot{q}}^{\;\; -}= \gamma^i_{q\dot{q}}
(u^{\underline{b}--}u^{\underline{c}i}- u^{\underline{c}--}
u^{\underline{b}i})$ and arrives at
\begin{eqnarray}\label{Ki=H+-i}
 & K^i =  \; {1\over 4} \; \hat{H}^{--\;++\; i}\; , \qquad \hat{H}^{--\;++\; i} := \,
\hat{H}_{\underline{a}\underline{b}\underline{c}}
u^{\underline{a}--} u^{\underline{b}++} u^{\underline{c}i}\; ,
\qquad
\end{eqnarray}

Thus we have completed the derivation of superstring equations of
motion from the superembedding equations.

\section{Superembedding description of AdS superstring}
\label{sec1.4}

\subsection{AdS superspace $AdSS^{(5,5|32)}$ as the solution of type IIB supergravity consraints}

The AdS superspace denoted by $AdSS^{(5,5|32)}$ (see
\cite{Howe+H=2000}) is the D=10 type IIB superspace the bosonic body
of which is $AdS_5\times S^5$. This is given by a solution of the
type IIB supergravity constraints (\ref{Ta=consSTR}),
(\ref{Tal1=str}), (\ref{Tal2=str}) with all but five form fluxes
equal to zero, this is to say
\begin{eqnarray}\label{H=0AdS}
H_{\underline{a}_1\underline{a}_2\underline{a}_3}=0\; , \qquad
R_{\underline{a}_1\underline{a}_2\underline{a}_3}=0\; , \qquad
C_0=0=\Phi\, . \qquad
\end{eqnarray}
The nonvanishing five form flux is characterized by a selfdual
constant tensor
\begin{eqnarray}\label{fAdS=sd}
& f^{\underline{a}_1\underline{a}_2\underline{a}_3\underline{a}_4\underline{a}_5}={1
\over 5!}
\epsilon^{\underline{a}_1\underline{a}_2\underline{a}_3\underline{a}_4\underline{a}_5
\underline{c}_1\underline{c}_2\underline{c}_3\underline{c}_4\underline{c}_5}
f_{\underline{c}_1\underline{c}_2\underline{c}_3\underline{c}_4\underline{c}_5}\,
,
 \quad \\ \label{df=0AdS}
 & df^{\underline{a}_1\underline{a}_2\underline{a}_3\underline{a}_4\underline{a}_5}=0\, . \qquad
\end{eqnarray}
The torsion and curvature two-forms of the $AdSS^{(5,5|32)}$ superspace are
expressed through this constant tensor and $\sigma$-matrices by (see
\cite{Kallosh:1998qs,AdSflat})
\begin{eqnarray}\label{AdSTa=}
  T^{\underline{a}} &=&  -i {\cal E} \wedge (I\otimes
       \sigma^{\underline{a}})
       {\cal E} \equiv  - i\; \left(
       E^{{\alpha}1}  \wedge
       E^{{\delta}1}+       E^{\underline{\alpha}2}  \wedge
       E^{{\delta}2}\right)\;
       \sigma^{\underline{a}}_{\underline{\alpha}\underline{\delta}}  \; ,
\\ T^{\underline{\; \alpha}}  &=& - {1\over R} E^{\underline{a}}
\wedge {\cal E}^{\underline{\beta}} \;
 f_{ \underline{a}\underline{b}_1\ldots
\underline{b}_4} (i\tau_2 \otimes \sigma^{\underline{b}_1\ldots
\underline{b}_4})_{\underline{\beta}} {}^{\underline{\alpha}}\; ,
\qquad \nonumber \\ \label{AdSTal=} && \qquad \Leftrightarrow \qquad
\begin{cases} T^{{\alpha}1}  = \; {1\over R} E^{\underline{a}}
\wedge E^{\underline{\beta}2} \;  f_{
\underline{a}\underline{b}_1\ldots \underline{b}_4}
(\sigma^{\underline{b}_1\ldots \underline{b}_4})_{{\beta}}
{}^{{\alpha}}\; , \cr T^{{\alpha}2}  =- {1\over R} E^{\underline{a}}
\wedge E^{\underline{\beta}1} \;  f_{
\underline{a}\underline{b}_1\ldots \underline{b}_4}
(\sigma^{\underline{b}_1\ldots \underline{b}_4})_{{\beta}}
{}^{{\alpha}}\; ,
\end{cases} \quad
\end{eqnarray}\begin{eqnarray}
\label{AdSRab=} R^{\underline{a}\underline{b}} &=&
 -{1\over 2R^2}  E^{\underline{a}} \wedge E^{\underline{b}} -
{4i\over R} {\cal E}^{\underline{\alpha}}  \wedge
       {\cal E}^{\underline{\delta}}
(i\tau_2 \otimes
{\sigma}_{\underline{c}_1\underline{c}_2\underline{c}_3})
{}_{\underline{\alpha}\underline{\delta}} f^{
\underline{a}\underline{b}\underline{c}_1\underline{c}_2\underline{c}_3}\; . \qquad \nonumber \\
&=&  -{1\over 2R^2}  E^{\underline{a}} \wedge E^{\underline{b}} +
{8i\over R} E^{\underline{\alpha}2}  \wedge
       E^{\underline{\delta}1}
f^{
\underline{a}\underline{b}\underline{c}_1\underline{c}_2\underline{c}_3}
{\sigma}_{\underline{c}_1\underline{c}_2\underline{c}_3}
{}_{{\alpha}{\delta}}\; . \qquad
\end{eqnarray}

To complete the definition of the $AdSS^{(5,5|32)}$ superspace we
should add that, in a suitable frame one can split the set of
bosonic supervielbein forms as $E^{\underline{a}}=(E^{\check{a}},
E^{\check{i}})\,,$ with $ {\check{a}}=0,1,\ldots , 4$ and
$\check{i}=1,\ldots , 5$ and find that the constant self-dual tensor
(\ref{fAdS=sd}) should have the form
\begin{eqnarray}\label{gframe}
f_{\check{a}_1\ldots\check{a}_5}=
\varepsilon_{\check{a}_1\ldots\check{a}_5}\; , \quad \qquad
f_{\check{i}_1\ldots \check{i}_5}=  \varepsilon_{\check{i}_1\ldots
\check{i}_5}\; ,
\end{eqnarray}
with all other components vanishing.
Then $ {\check{a}}=0,1,\ldots , 4$ is identified as the vector index
of the 5d space tangent to $AdS_5\,,$ and $i=1,\ldots , 5$ -- as the
vector index of the space tangent  to  $S^5$.  The constant $R$ in
(\ref{AdSTal=}), (\ref{AdSRab=}) defines the radius of $AdS_5$ or
$S^5$ (these radii are equal).

Now the superembedding description of the AdS superstring can be
obtained by specializing the equations describing superstring in
general supergravity background to a particular form of this
background given by Eqs. (\ref{AdSTa=}), (\ref{AdSTal=}),
(\ref{AdSRab=}), (\ref{fAdS=sd}), (\ref{df=0AdS}). But before
turning to this, we describe some properties of constant selfdual
tensor
$f^{\underline{a}_1\underline{a}_2\underline{a}_3\underline{a}_4
\underline{a}_5}:= f^{\underline{[5]}}=\, {1\over
5!}\epsilon^{[5][5']}f_{[5']}$ as they are seen in stringy spinor
moving frame.

\subsection{Constant five form flux in stringy moving frame}

Below we will mainly use a seemingly SO(1,9) covariant description
by Eqs.  (\ref{AdSTa=}), (\ref{AdSTal=}), (\ref{AdSRab=}) and
(\ref{fAdS=sd}), (\ref{df=0AdS}) so that a big part of our results
are applicable for superstring in a generic constant 5-form flux
background.

Although the distinction between self-duality and anti-self
duality is conventional, the selfduality of the constant flux (\ref{fAdS=sd}) is singled out by
that we use the $sigma$--matrix representation with self-dual
$\sigma_{\alpha\beta}^{[5]}:=\sigma^{\underline{a}_1\underline{a}_2\underline{a}_3\underline{a}_4
\underline{a}_5}_{\alpha\beta}$
(which implies  anti-selfdual $\tilde{\sigma}^{[5]\; \alpha\beta}:=
\tilde{\sigma}^{\underline{a}_1\underline{a}_2\underline{a}_3\underline{a}_4
\underline{a}_5\; \alpha\beta}$),
\begin{eqnarray}\label{s5=es5}
\sigma^{[5]}_{\alpha\beta}={1\over 5!} \epsilon^{[5][5^\prime]}
\sigma_{[5^\prime]\;\alpha\beta}\; , \qquad \tilde{\sigma}^{[5]\;
\alpha\beta}=- {1\over 5!} \epsilon^{[5][5^\prime]}
\tilde{\sigma}_{[5^\prime]}^{\alpha\beta}\; . \qquad
\end{eqnarray}

It is convenient to introduce the spinor moving frame components of
the constant selfdual tensor (which are not obliged to be constant
but rather depend on the coordinate of the superstring worldsheet
superspace),
\begin{eqnarray}\label{f--++ijk:=}
f^{ijk}:=   & f^{--\, ++\, i_1i_2i_3} = f^{
\underline{c}_1\underline{c}_2\underline{c}_3\underline{c}_4\underline{c}_5}u^{--}_{
\underline{c}_1}u^{++}_{\underline{c}_2}
u^{i_1}_{\underline{c}_3}u^{i_2}_{\underline{c}_4}u^{i_3}_{\underline{c}_5}\; , \qquad \nonumber  \\
& f^{i_1i_2i_3i_4i_5} = f^{
\underline{c}_1\underline{c}_2\underline{c}_3\underline{c}_4\underline{c}_5}u^{i_1}_{
\underline{c}_1}u^{i_2}_{\underline{c}_2}
u^{i_3}_{\underline{c}_3}u^{i_4}_{\underline{c}_4}u^{i_5}_{\underline{c}_5}\; , \qquad \nonumber  \\
& f^{\pm\pm\, i_1i_2i_3i_4} = f^{
\underline{c}_1\underline{c}_2\underline{c}_3\underline{c}_4\underline{c}_5}u^{\pm\pm}_{
\underline{c}_1}u^{i_1}_{\underline{c}_2}
u^{i_2}_{\underline{c}_3}u^{i_3}_{\underline{c}_4}u^{i_4}_{\underline{c}_5}\;
. \qquad
\end{eqnarray}
The self-dulaity equation (\ref{fAdS=sd}) implies
\begin{eqnarray}\label{f5--=id3}
f^{ijk}:=   f^{--\, ++\, ijk} = & \;  {1\over
5!}\epsilon^{ijkl_1l_2l_3l_4l_5}f^{l_1l_2l_3l_4l_5}\, ,
\qquad \nonumber \\   f^{++ \, ijkl}= &- {1\over 4!}\epsilon^{ijkli'j'k'l'}f^{++\, i'j'k'l'}\,  ,  \qquad \nonumber \\
 f^{-- \, ijkl}= &\; {1\over 4!}\epsilon^{ijkli'j'k'l'}f^{--\, i'j'k'l'}\,  ,  \qquad
\end{eqnarray}
Now one can prove that, as a consequence of (\ref{fAdS=sd}) and of
the properties (\ref{v+sv+=u++}), (\ref{v-sv-=u--}), (\ref{uIu=I})
of stringy harmonics, the following identities hold
\begin{eqnarray}\label{f5--=id}
f^{--\underline{a}\underline{b}\underline{c}\underline{d}}
(v^-_{{q}}\sigma_{\underline{a}\underline{b}\underline{c}\underline{d}})^{\alpha}=0
\; , \qquad
f^{++\underline{a}\underline{b}\underline{c}\underline{d}}
(v^+_{\dot{q}}\sigma_{\underline{a}\underline{b}\underline{c}\underline{d}})^{\alpha}=0
\; ,
\end{eqnarray}
where
\begin{eqnarray}\label{f5--:=}
f^{\pm\pm\;
\underline{b}_1\underline{b}_2\underline{b}_3\underline{b}_4}=
u^{\pm\pm}_{\underline{a}}f^{\underline{a}\underline{b}_1\underline{b}_2\underline{b}_3\underline{b}_4}
\; .
\end{eqnarray}
Notice by pass that tensors $f^{\pm\pm\;
\underline{b}_1\underline{b}_2\underline{b}_3\underline{b}_4}$
characterize the movement of superstring. For instance, when
superstring moves in the $AdS$ part of superspace and is frozen to a
point on  $S^5$, one can chose the frame in such a way that
$u_{\underline{a}}^{\pm\pm}=\delta_{\underline{a}}{}^{\check{c}}u_{\check{c}}^{\pm\pm}$
and, then $f^{\pm\pm\;
\underline{c}_1\underline{c}_2\underline{c}_3\underline{c}_4}=\,
u_{\check{a}}^{\pm\pm}\,
\epsilon^{\check{a}\check{b}_1\check{b}_2\check{b}_3\check{b}_4} \,
\delta_{\check{b}_1}{}^{\underline{c}_1}\delta_{\check{b}_2}{}^{\underline{c}_2}
\delta_{\check{b}_3}{}^{\underline{c}_3}
\delta_{\check{b}_4}{}^{\underline{c}_4}$.

 To prove (\ref{f5--=id}), we observe  that
 (\ref{fAdS=sd}) and (\ref{s5=es5}) imply
$f_{[5]}\sigma^{[5]}_{\alpha\beta}=0$ (but
$f_{[5]}\tilde{\sigma}^{[5]\; \alpha\beta}\not=0$) which implies
$f^{\underline{a}\underline{b}\underline{c}\underline{d}\underline{e}}
\sigma_{\underline{b}\underline{c}\underline{d}\underline{e}}{}_\beta{}^{\alpha}={1\over
10} \sigma^{\underline{a}}_{\beta\gamma}f_{[5]}\tilde{\sigma}^{[5]\;
\gamma\alpha}$.  Multiplying this by $u_{\underline{a}}^{--}
v^{-\beta}_q$ and $u_{\underline{a}}^{++} v^{+\beta}_{\dot{q}}$ one
finds that (\ref{f5--=id}) holds due to the identities
$u_{\underline{a}}^{--} (v^{-}_q\sigma^{\underline{a}})_\gamma=0$
and $u_{\underline{a}}^{++}
(v^{+}_{\dot{q}}\sigma^{\underline{a}})_\gamma=0$, respectively.

Nitice that in such a way we cannot prove
 \begin{eqnarray}\label{f5--=id4}
f^{--\underline{a}\underline{b}\underline{c}\underline{d}}
(\sigma_{\underline{a}\underline{b}\underline{c}\underline{d}}v^-_{\dot{q}})_{\alpha}=0\; , \qquad
f^{++\underline{a}\underline{b}\underline{c}\underline{d}}
(\sigma_{\underline{a}\underline{b}\underline{c}\underline{d}}v^+_q)^{\alpha}=0\; . \qquad
\end{eqnarray}
To show that these are true, let us observe first that, due to
(\ref{f5--=id}),  we need only to prove the vanishing of their
contractions with, respectively, $v_{\dot{q}}^{+\alpha}$ and
$v_{{q}}^{-\alpha}$. For these one finds that
\begin{eqnarray}\label{f5--=id6}
f^{--\underline{a}\underline{b}\underline{c}\underline{d}}
(v^+_{\dot{p}}\sigma_{\underline{a}\underline{b}\underline{c}\underline{d}}v^-_{\dot{q}})=f^{--i_1i_2i_3i_4}
\tilde{\gamma}^{i_1i_2i_3i_4}_{\dot{p}\dot{q}}= 0
\; , \qquad \nonumber \\
f^{++\underline{a}\underline{b}\underline{c}\underline{d}}
(v^-_{{p}}\sigma_{\underline{a}\underline{b}\underline{c}\underline{d}}v^+_{{q}})=
f^{++i_1i_2i_3i_4} \tilde{\gamma}^{i_1i_2i_3i_4}_{{p}{q}}=0
\; , \qquad
\end{eqnarray}
where the last equalities follow from (\ref{f--++ijk:=}), (\ref{f5--=id3}) and the
anti-self duality (self-duality) of $\tilde{\gamma}^{i_1i_2i_3i_4}$
(${\gamma}^{i_1i_2i_3i_4}$),
\begin{eqnarray} \tilde{\gamma}_{\dot{q}\dot{p}}^{ijkl}= -
& {1\over 4!}
\epsilon^{ijkli'j'k'l'}\tilde{\gamma}_{\dot{q}\dot{p}}^{i'j'k'l'}\;
, \qquad {\gamma}^{ijkl}_{qp}= + {1\over 4!}
\epsilon^{ijkli'j'k'l'}{\gamma}_{qp}^{i'j'k'l'}
\end{eqnarray} which
correspond to the duality properties (\ref{s5=es5}) of the 10D
$\sigma$ matrices.

To conclude, the only nonvanishing contribution to the expressions
$f^{\pm\pm\underline{[4]}} \sigma_{\underline{[4]}}=
f^{\pm\pm\underline{a}\underline{b}\underline{c}\underline{d}}
(\sigma_{\underline{a}\underline{b}\underline{c}\underline{d}})_{\beta}{}^{\alpha}$
is
\begin{eqnarray}\label{f5--=id7}
f^{--\underline{a}\underline{b}\underline{c}\underline{d}}
(v^+_{\dot{q}}\sigma_{\underline{a}\underline{b}\underline{c}\underline{d}}v^+_q)
&= & f^{++\underline{a}\underline{b}\underline{c}\underline{d}}
(v^-_{{q}}\sigma_{\underline{a}\underline{b}\underline{c}\underline{d}}v^-_{\dot{q}})=
4 f^{ijk} \gamma^{ijk}_{q\dot{q}} \; . \qquad
\end{eqnarray}

\subsection{Worldsheet superspace geometry of the superstring in AdS superspace}

The pull--backs of the AdS supervielbein on the worldsheet
superspace have the form of (\ref{hEa=STR}) and (\ref{hEf1f2=STR}),
\begin{eqnarray}\label{hEa=AdS}
& \hat{E}^{\underline{a}} = {1\over 2}e^{++} u^{--\underline{a}} +
{1\over 2}e^{--} u^{++\underline{a}} \; ,  \qquad
\\ \label{hEf1f2=AdS}
 & \hat{E}^{\alpha 1} = e^{+q} v^{-\alpha}_{q} + e^{++}
\chi^{-\dot{q}}_{_{++}}  v^{+\alpha}_{\dot{q}}\; , \qquad
\hat{E}^{\alpha 2} = e^{-\dot{q}} v^{+\alpha}_{\dot{q}}  + e^{--}
\chi^{+{q}}_{_{--}}  v^{-\alpha}_{q}\; .  \qquad
\end{eqnarray}
As far as the $H_3$ flux in AdS superspace is equal to zero, Eq.
(\ref{H=0AdS}), the string equations of motion (\ref{Ki=H+-i}) read
\begin{eqnarray} \label{Ki=0AdS} &  K^i =0\; , \qquad
\end{eqnarray}
so that the generalized Cartan form, which provide the
supersymmetric generalization of the second fundamental form for the
AdS superstring worldvolume superspace, are given by Eqs.
(\ref{Om--iSTR=}), (\ref{Om++iSTR=})
\begin{eqnarray}\label{Om--i=AdS}
\Omega^{\!^{--i}} = - 4i e^{+q} (\gamma^i \chi_{\!_{++}}^{-})_q +
e^{\!^{++}} \Omega^{^{--i}}_{\!_{++}} \; , \qquad \nonumber
\\  \Omega^{\!^{++i}}= - 4i e^{-\dot{q}} (\chi_{\!_{--}}^{+}\gamma^i)
_{\dot{q}} + e^{\!^{--}} \Omega^{^{++i}}_{\!_{--}}  \; . \qquad
\end{eqnarray}
 The worldvolume superspace torsion forms read (see  Eqs. (\ref{De+-=STR}), (\ref{De-dq=STR}))
\begin{eqnarray}\label{De++=AdS}
& De^{\!^{++}}=-2i e^{+q}\wedge e^{+q} \; , \qquad  De^{\!^{--}}=-2i
e^{-\dot{q}}\wedge e^{-\dot{q}} \; , 
  \qquad
\\ \label{De+q=AdS} De^{+q} &= -e^{++}\wedge e^{-\dot{p}} \left(2i
(\chi_{--}^{\quad +}\gamma^{i})_{\dot{p}}\,
({\gamma}^{i}\chi_{++}^{\quad -})_q  + {2\over R}\,
 f^{ijk} \gamma^{ijk}_{q\dot{q}}\right)
+ \qquad   \nonumber \\ & +{1\over 2}\, e^{++}\wedge e^{--}
\Omega_{--}^{\;
++\, i }({\gamma}^{i}\chi_{++}^{\quad -})_q  \; , \qquad \\
\label{De-dq=AdS} De^{-\dot{q}}&= -2i e^{--}\wedge e^{+{p}}
\left(2i (\gamma^{i}\chi_{++}^{\quad -})_{p}\, (\chi_{--}^{\quad
+}{\gamma}^{i})_{\dot{q}} + {2\over R} \,
 f^{ijk} \gamma^{ijk}_{q\dot{q}}\right) - \nonumber \qquad  \\
 & -{1\over 2}\, e^{++}\wedge e^{--} \Omega_{++}^{\; --\, i
}(\chi_{--}^{\quad +}{\gamma}^{i})_{\dot{q}} \; \,  \qquad
\end{eqnarray}
To obtain these expressions the identities (\ref{f5--=id}) have to
be used; they are also needed to find  that
$D_{--}\chi_{++}{}^-_{\dot{q}}=D_{-\dot{p}}\chi_{++}{}^-_{\dot{q}}=0$
and $D_{++}\chi_{--}{}^+_{{q}}=D_{+p}\chi_{--}{}^+_{{q}}=0$ which
can be summarized by
\begin{eqnarray}\label{Dchi++=AdS}
 D\chi_{++}{}^-_{\dot{q}}& =  -{1\over 2} e^{+q}\gamma^{i}_{q\dot{q}}
\Omega_{++}^{\; --\, i} + e^{++}D_{++}\chi_{++}{}^-_{\dot{q}}\; , \qquad \\
\label{Dchi--=AdS}
 D\chi_{--}{}^+_{{q}}& =
-  {1\over 2} e^{-\dot{p}}
 \gamma^{i}_{q\dot{q}}
\Omega_{--}^{\; ++\, i} +e^{--}D_{--}\chi_{--}{}^+_{{q}}\; . \qquad
\end{eqnarray}

The pull--back of the Riemann curvature two form of the AdSS
superspace to the worldvolume superspace reads
\begin{eqnarray}\label{hR=AdS}
& \hat{R}^{\underline{a}\underline{b}} =
 - {1\over 2}e^{++}\wedge e^{--}  \left({1\over R^2} u^{-- [\underline{a}} \wedge
 u^{\underline{b}]++} + {16i\over R}
 f^{\underline{a}\underline{b}ijk}(\chi_{--}^{\;\; +} \gamma^{ijk}\chi_{++}^{\;\;
-})\right)
 - \qquad \nonumber \\ & -
{24i\over R} \, e^{--}\wedge e^{+{q}}\, f^{--\,
\underline{a}\underline{b}ij} (\gamma^{ij}\chi_{--}^{\;\; +})_q +
{24i\over R} \, e^{++}\wedge e^{-\dot{q}}\, f^{++\,
\underline{a}\underline{b}ij} (\tilde{\gamma}^{ij}\chi_{++}^{\;\;
-})_{\dot{q}} -  \nonumber \\ & - {8i\over R}\, e^{+q}\wedge
e^{-\dot{q}} f^{ \underline{a}\underline{b}ijk}
{\gamma}^{ijk}_{q\dot{q}}\; . \qquad
\end{eqnarray}
In its derivation we have used the following consequences of the
properties (\ref{u++=v+v+2}), (\ref{u--=v-v-2}), (\ref{ui=v+v-2})
of spinor moving frame variables (\ref{VSTRin})
\begin{eqnarray}\label{fvs3v=}
  f^{\underline{a}\underline{b}\underline{[3]}}
v^-_p\sigma_{\underline{[3]}}v^-_{{q}} &=& - 3f^{--\,
\underline{a}\underline{b}ij}\gamma^{ij}_{pq} \qquad
f^{\underline{a}\underline{b}\underline{[3]}}
v^+_{\dot{p}}\sigma_{\underline{[3]}}v^+_{\dot{q}}=- 3f^{++\,
\underline{a}\underline{b}ij}\tilde{\gamma}^{ij}_{\dot{p}\dot{q}}
 \; , \qquad \nonumber \\ &&  f^{\underline{a}\underline{b}\underline{[3]}}
v^-_q\sigma_{\underline{[3]}}v^+_{\dot{q}} =
f^{\underline{a}\underline{b}ijk}\gamma^{ijk}_{q\dot{q}}\; . \qquad
\end{eqnarray}

Substituting  Eqs. (\ref{hR=AdS}) and (\ref{Om--i=AdS}) into Gauss
and Ricci equations (\ref{DOm0=STR}) and (\ref{DOmij=STR}), one
finds the following expressions for the 2d Riemann curvature two
form
\begin{eqnarray}
 \label{DOm0=AdS}  d\Omega^{(0)} \;
&=  e^{++}\wedge e^{--} \left({1\over 4}\Omega_{++}^{\; -- i}\Omega_{--}^{\;
++ i}+ {2i\over R} f^{ijk}(\chi_{--}^{\;\; +}
\gamma^{ijk}\chi_{++}^{\;\; -})\right)
 + \qquad \nonumber \\ & +i\, e^{--}\wedge e^{+{q}}\, \Omega_{--}^{\; ++ i}
(\gamma^{i}\chi_{++}^{\;\; -})_q  - i \, e^{++}\wedge
e^{-\dot{q}}\, \Omega_{++}^{\; -- i} (\chi_{++}^{\;\;
-}{\gamma}^{i})_{\dot{q}} + \nonumber
\\ & + e^{+q}\wedge e^{-\dot{q}} \left(
4(\gamma^{i}\chi_{++}^{\;\; -})_q (\chi_{++}^{\;\;
-}{\gamma}^{i})_{\dot{q}} + {2i\over R}\, f^{ijk}
{\gamma}^{ijk}_{q\dot{q}}  \right)\;  \qquad
\end{eqnarray}
and for the curvature of the normal bundle
\begin{eqnarray}
\label{DOmij=AdS} \mathbb{R}^{ij} \; &= e^{++}\wedge e^{--}
\left(\Omega_{--}^{\; ++ [i}\Omega_{++}^{\; -- j]}- {8i\over R}
f^{ijk_1k_2k_3}(\chi_{--}^{\;\; +} \gamma^{k_1k_2k_3}\chi_{++}^{\;\;
-})\right)
 + \qquad \nonumber \\ & +4i\, e^{--}\wedge e^{+{q}}\, \left(\Omega_{--}^{\; ++ [i}
(\gamma^{j]}\chi_{++}^{\;\; -})_q - {8\over R} f^{--
ijkl}(\gamma^{kl}\chi_{--}^{\;\; +})_q \right) - \qquad \nonumber \\
&- 4i \, e^{++}\wedge e^{-\dot{q}}\, \left(\Omega_{++}^{\; -- [i}
(\chi_{++}^{\;\; -}{\gamma}^{j]})_{\dot{q}}+ {8\over R} f^{++
ijkl}(\tilde{\gamma}^{kl}\chi_{++}^{\;\; -})_{\dot{q}} \right) + \qquad
\nonumber
\\ & -8 e^{+q}\wedge e^{-\dot{q}} \left(
2(\gamma^{[i}\chi_{++}^{\;\; -})_q (\chi_{++}^{\;\;
-}{\gamma}^{j]})_{\dot{q}} + {i\over R}\, f^{ijk_1k_2k_3}
{\gamma}^{k_1k_2k_3}_{q\dot{q}}  \right)\; . \qquad
\end{eqnarray}
The Peterson-Codazzi equations (\ref{DOm+-i=STR}) for superstring in $AdS_5\times S^5$ superspace read
\begin{eqnarray}\label{DOm--i=AdS}
 &D\Omega^{--\, i} = \hat{R}^{--\, i}=
- {8i\over R} e^{++}\wedge e^{--}f^{--ijkl}(\chi_{--}^{\;\; +}
\gamma^{jkl}\chi_{++}^{\;\; -})
 - \qquad  \\ \nonumber
&-{24i\over R}\, e^{++}\wedge e^{-\dot{q}}\,  f^{
ijk}(\tilde{\gamma}^{jk}\chi_{++}^{\;\; -})_{\dot{q}}  - {8i\over R}
e^{+q}\wedge e^{-\dot{q}}  \, f^{--ijkl}
{\gamma}^{jkl}_{q\dot{q}} \; , \; \\
\label{DOm++i=AdS}
 &D\Omega^{++\, i} =\hat{R}^{++\, i}=
- {8i\over R} e^{++}\wedge e^{--}f^{++ijkl}(\chi_{--}^{\;\; +}
\gamma^{jkl}\chi_{++}^{\;\; -})
 - \qquad \\ \nonumber  & -{24i\over R} e^{--}\wedge e^{+{q}}\, f^{
ijk}(\gamma^{jk}\chi_{--}^{\;\; +})_q   - {8i\over R} e^{+q}\wedge
e^{-\dot{q}}  \, f^{++ijkl} {\gamma}^{jkl}_{q\dot{q}}  \; . \;
\end{eqnarray}
These  do not give us new
information after Eqs. (\ref{Dchi++=AdS}) and (\ref{Dchi--=AdS}) are
taken into account.

Thus we have completed the
construction of superembedding approach description of the
superstring in $AdSS^{(5,5|32)}$ superspace.

\section{Conclusion}

In this contribution we have presented a brief review of spinor
moving frame formulation, generalized action principle and
superembedding approach to super-$p$-branes and (in Sec.
\ref{sec1.3})  have elaborated in detail the superembedding approach
to superstring in general type IIB supergravity background. On this
basis we have given (in Sec. \ref{sec1.4}) the complete
superembedding description of superstring in $AdS_5\times S^5$
superspace. To our best knowledge, such a description of AdS
superstring has not been developed before and we hope that it will
be helpful in searching for new exact solutions of the AdS
superstring equations which, in its turn, might be useful for
further study and applications of the AdS/CFT correspondence.

\medskip

{\bf Acknowledgments.\, }
The author thanks Jos\'e de Azc\'arraga and Dmitri Sorokin for
discussions, useful comments and collaboration on early stages of
this work which has been supported by the Basque Foundation for
Science, {\it Ikerbasque}, and partially by research grants from the
Spanish MICI (FIS2008-1980), the INTAS (2006-7928) as well as by the Ukrainian
Academy of Sciences and Russian RFFI grant 38/50--2008.


\end{document}